# Underwater dual-magnification imaging for automated lake plankton monitoring

Ewa Merz[1]*, Thea Kozakiewicz[1]*, Marta Reyes[1]*, Christian Ebi[1]*, Peter Isles[1]*, Marco Baity-Jesi[1]*, Paul Roberts[2,3], Jules S. Jaffe[2], Stuart Dennis[1], Thomas Hardeman[1], Nelson Stevens[1], Tom Lorimer[1,2], Francesco Pomati[1]*

* these authors contributed equally

[1]Swiss Federal Institute of Aquatic Science and Technology (Eawag), Ueberlandstrasse 133, 8600 Dübendorf, Switzerland

[2]Scripps Institution of Oceanography, University of California San Diego,9500 Gilman Drive, La Jolla, CA 92093-0238, United States

[3]Monterey Bay Aquarium Research Institute (MBARI), 7700 Sandholdt Road, Moss Landing, CA 95039, United States

Corresponding authors: Ewa Merz (ewa.merz@eawag.ch), Francesco Pomati (francesco.pomati@eawag.ch)

# Abstract

We present an approach for automated *in-situ* monitoring of phytoplankton and zooplankton communities based on a dual magnification dark-field imaging microscope/camera. We describe the Dual Scripps Plankton Camera (DSPC) system and associated image processing, and assess its capabilities in detecting and characterizing plankton species of different size and taxonomic categories, and in measuring their abundances in both laboratory and field applications. In the laboratory, body size and abundance estimates by the DSPC significantly and robustly scale with the same measurements derived by traditional microscopy. In the field, a DSPC installed permanently at 3 m depth in Lake Greifensee (Switzerland), delivered images of plankton individuals, colonies, and heterospecific aggregates without disrupting natural arrangements of interacting organisms, their microenvironment or their behavior at hourly timescales. The DSPC was able to track the dynamics of taxa in the size range between ~10 μm to ~ 1 cm, covering virtually all the components of the planktonic food web (including parasites and potentially toxic cyanobacteria). Comparing data from the field-deployed DSPC to traditional sampling and microscopy revealed a general overall agreement in estimates of plankton diversity and abundances, despite imaging limitations in detecting small phytoplankton species and rare and large zooplankton taxa (e.g. carnivorous zooplankton). The most significant disagreements between traditional methods and the DSPC resided in the measurements of community properties of zooplankton, organisms that are heterogeneously distributed spatially and temporally, and whose demography appeared to be better captured by automated imaging. Time series collected by the DSPC depicted ecological succession patterns, algal bloom dynamics and circadian fluctuations with a temporal frequency and morphological resolution that would have been impossible with traditional methods. We conclude that the DSPC approach is suitable for stable long-term deployments, and robust for both research and water quality monitoring. Access to high frequency, reproducible and real-time data of a large spectrum of the planktonic ecosystem might represent a breakthrough in both applied and fundamental plankton ecology.

0p5x: 0.5 times magnification

5p0x: 5.0 times magnification

CNN: convolutional neural network

# 1. Introduction

Plankton are a key component of all water bodies and the Earth's biosphere, being crucial for important ecosystem processes such as carbon and nutrient cycling, and provide essential services to human society (e.g. clean water and fisheries) (Behrenfeld et al., 2001; Falkowski, 2012). The combination of relatively short lifespans and strong sensitivity to environmental conditions makes planktonic organisms effective indicators of environmental change and ecosystem health, and are used to assess water and aquatic ecosystem quality worldwide (Xu, 2001; Directive 2000/60/EC). Information about plankton biomass and community composition, including variation among key functional traits, is essential to assess the state of ecological systems, their resilience to change, extinctions and invasions, and to manage the ecosystem services that they provide (Bartley et al., 2019; Cardinale et al., 2012; Pomati et al., 2011). Assessing plankton community composition is critically important for the monitoring and forecasting of harmful algal blooms (events of algal overgrowth that are dominated by toxic species). In freshwaters, cyanobacterial blooms cause large ecological and economical damage (Huisman et al., 2018), and have been increasing worldwide due to effects of land and climate change (Chorus and Bartram, 1999; Ho et al., 2019).

Given the importance of plankton for ecosystem processes and services, there is not only scientific interest, but also a societal need and a policy requirement for monitoring plankton communities at low cost and ideally in real-time. The assessment of plankton community composition and taxa relative abundance is however the most difficult, time consuming and expensive aspect (Pomati et al., 2011). Dedicated laboratories use skilled and trained taxonomists to identify and count microalgae and zooplankton manually, sometimes taking several hours to count one sample. Taxonomists frequently misclassify items, count some objects more than once or overlook others (MacLeod et al., 2010) and mistakes are not traceable back in time. In many monitoring programs, microscope counts are only conducted weeks or months after samples are collected, making this approach unable to provide information in near real-time. Sample collection, transport and storage also demand proper infrastructure and personnel time. Aquatic ecosystem monitoring would hence tremendously benefit from an automated, reliable, standalone system for plankton identification, classification and counting that produces transparent and reproducible results (not dependent on personal judgement), and that does not require sample collection and storage (additional costs).

Several automated quantitative tools for plankton counting have been established: laboratory-based (e.g. Flow-cam, flow-cytometry, microscopy-imaging, Zooscan) and field-based (e.g. fluorescence probes, Cytobuoy, Flowcytobot, imaging cameras) (Lombard et al., 2019). Laboratory-based tools can automate counting and sorting, but still require sample collection and processing, which changes the nature of the sample and requires time, resources and infrastructure. Field applications are more promising, but they have their own limitations. Fluorescence sensors measuring chlorophyll-a, phycocyanin, or other photosynthetic pigments are common and widely deployed in the field, but they are unable to resolve taxonomic distinctions beyond coarse levels, and pigment content within and across species can vary widely, making estimates of abundances and biomass difficult (Johnson and Martiny, 2015). Automated in situ flow-cytometry has shown encouraging results for high frequency monitoring of phytoplankton community dynamics and morpho-physiological traits (Fontana et al., 2018; Hunter-Cevera et al., 2016; Pomati et al., 2011; Sosik et al., 2010). Flow-cytometry complemented by water physics and chemistry data, and machine learning for analysis, demonstrated the potential of high-frequency information for modelling environmental responses and predicting phytoplankton dynamics (Thomas et al., 2018). Dynamics of single species of phytoplankton and microzooplankton can be tracked and studied with flow-cytometry if assisted by imaging (Hunter-Cevera et al., 2016; Lombard et al., 2019). Still, flow-cytometers require constant maintenance and calibration, and may disrupt natural spatial aggregates while sampling. Additionally, they have limited dynamic range in terms of size, often being able to follow the dynamics of a single trophic level only (Lombard et al., 2019). For phytoplankton, particularly, high frequency data of herbivore zooplankton could be essential to study variation in top-down controls as a response to environmental change (Murphy et al., 2020).

Here we describe a monitoring approach based on dark-field underwater imaging, by means of a dual-magnification camera derived from the Scripps Plankton Camera (SPC) system (Orenstein et al., 2020). It was designed to be a flexible, easily configurable imaging system that can observe objects from 10s of microns to several millimeters, and sample at high temporal frequency, with minimal influence on the fluid being imaged. Similar cameras have been successfully deployed in marine systems (Campbell et al., 2020; Kenitz et al., 2020; Orenstein et al., 2020), but never deployed before in freshwater environments. In this study, we compared the DSPC to traditional microscopy and high-frequency fluorescence-based monitoring using both field and laboratory comparisons. We then used this tool to observe plankton dynamics and trait distributions (e.g. size) across the whole planktonic food web *in-situ*, with high temporal resolution. Our results promote underwater imaging as an auspicious approach to generate empirical high-resolution plankton time series.

# 2. Methods

## 2.1 Description of the dual magnification camera

### 2.1.1 Instrument's design

The Plankton Camera (DSPC) used in this study is a dual-magnification darkfield underwater microscope, based on the Scripps Plankton Camera (SPC) system (Orenstein et al. 2020). When compared to the original SPC, the DSPC has several new features designed specifically to target plankton monitoring in freshwaters. First, it is a dual-magnification microscope, extending the size range that can be photographed by overlapping spatial resolution (**Fig. 1**). Based on our trials, the highest magnification (5p0x) has a detection range between ~ 10 µm - 150 µm, while the lower magnification (0p5x) ranges between ~ 100 µm - 1 cm. These estimates depend on the features of the imaged objects (e.g. colour, transparency) and therefore are not fixed. The imaged volumes per frame for the two magnifications are (from high resolution to blob detection): 5p0x = 0.2 - 10 µL, 0p5x = 4 - 200 µL, respectively. Second, internal batteries allow autonomous measurements for up to 3 hours without recharging, which is needed for manual depth profiling or sampling of remote sites (i.e. lakes, ponds). Finally, the distance between light sources and objective has been reduced to 5 cm to diminish illumination path length, and allow transmission of light in turbid waters (e.g. during thick algal blooms).

The DSPC is relatively affordable and compact enough for one person to carry (57 cm L x 19 W, ~ 27.2 kg weight). The instrument is composed of two housings, one with the LED light sources and another with the controlling hardware, cameras and lenses (**Fig. 1**). The instrument has no moving parts; therefore, it has low maintenance requirements (only cleaning of external surfaces). Housings have facing acrylic viewports that allow transmitting light from the LED to the microscopes. Water flows freely between the two viewports. To reduce fouling, our system was complemented with a UV lamp (https://amloceanographic.com/biofouling-control/) installed facing the acrylic viewports (**Fig. 1**). The on-board computer uses a Linux based operating system. The camera works automatically, there is no need for regular calibration and, if installed in a monitoring platform like the DSPC at Lake Greifensee, pictures can be available in real time, if power supply, computer and internet are available. Acquisition rate for pictures can be easily modified and the maximum is 10 processed frames per second (recording of videos is also possible). The actual specifics of the instrument can be found in **Supplemental Table S1**. The configurations of the machine used for imaging are reported in the Methods section relative to laboratory and field application.

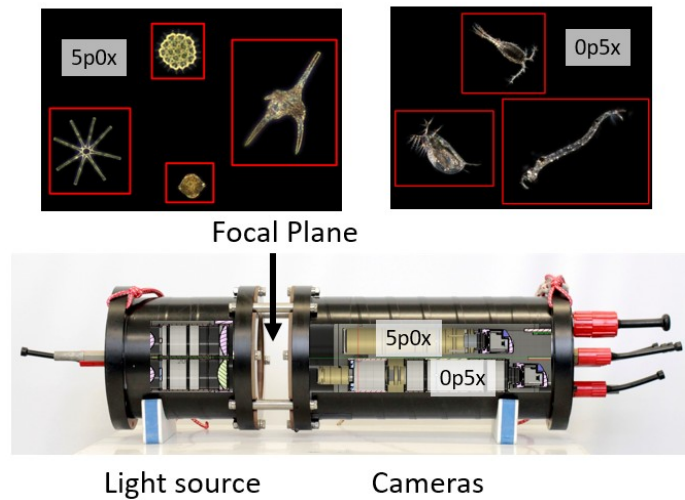

**Fig. 1. Structural drawing of the dual-magnification plankton camera, showing LED light source points, the two cameras with different magnification.** The focal plane is where water flows through and objects are imaged. The instrument captures full frame pictures, and automatically performs object detection (red boundaries around objects), cuts out these regions of interest (ROIs) that are stored, while full frames are usually discarded (Orenstein et al., 2020). Objects depicted as ROIs are only for display (not to scale).

## 2.1.2 Image processing

We use a modified version of the original Python SPCConvert image processing script (Orenstein et al., 2020), which converts raw regions of interest (ROIs) from the plankton camera into image mosaics and extracts statistics per processed sample and features of detected objects. The process consists of color conversion, edge-detection and segmentation, morphological feature extraction, foreground masking and inverse filtering of masked foreground. The results are then saved into a new directory and a web app (html file) is built from the data showing converted pictures. The program creates a spreadsheet with 64 features (e.g. area, aspect ratio, max. length or orientation, color features, and more) of every detected object. The Python code can be found here:
https://github.com/tooploox/SPCConvert.

Because of their complex shape and moveable appendages, measuring body size in zooplankton is not straightforward: often, zooplankton can be much wider than they are long when antennas are included (e.g. copepods), meaning that the major axis cannot reliably be used. Moreover, the presence of predators and other stressors also influences the overall shape of zooplankton, and can induce the growth of defensive structures (e.g. spines, helmets, and elongated tails) further complicating estimation of body size (Tollrian and Drew Harvell, 1999). To overcome these complexities, we developed The Daphnia Ruler (https://github.com/nelstevens/The-Daphnia-ruler), which effectively removes these appendages from calculations of zooplankton body size. Using the same edge detection concepts as SPCconvert, the outline of the zooplankton is refined by repeatedly eroding and dilating the binary image with round structuring elements of varying sizes, until a threshold solidity is reached indicating that all protuberances have been sufficiently removed. The resulting ellipse can then be treated as the core body and measured as for any other spheroid planktonic object.

## 2.2. Annotation of images and classification based on deep learning

A large number of labelled pictures is needed to create a training database for an automatic classification based on deep learning. We utilised a free and open source platform, Taxonify, which expedites the process of species annotation. Through this platform, one can rapidly label pictures to taxon level, and can add additional attributes to the pictures, i.e colonies, dividing, and parasites present. More information can be found on https://www.taxonify.org, and software codes are freely available at https://github.com/tooploox/taxonify_gui. For this study, we annotated a variable amount of images per class of interest (mainly zooplankton, in the 0p5x magnification), ranging from a minimum of 10 (*Chaoborus*) to a maximum of 3321 (*Dinobryon*) (**Supplemental Table S2**). We then developed an automated deep learning classifier for objects detected by the 0p5x magnification of the DSPC. The model we used was a convolutional neural network (CNN), trained to distinguish 34 different classes of zooplankton, phytoplankton and other suspended solids (junk) (**Supplemental Table S2**). In the SI text we describe the used CNN and its hyperparameters, and we show a series of validation benchmarks. We provide access to the code and a ready-to-use release of the classifier, at the following address: https://github.com/mbaityje/plankifier/releases/tag/v1.1.1. The default code settings contain all hyperparameter values used in this study.

## 2.3 Laboratory calibration of DSPC imaging

To use the DSPC in the laboratory, the space between the two viewports (**Fig. 1**) was covered with a black cloth and flooded with water. Tissue culture flasks containing the imaged samples were manually held in front of the focal plane of the DSPC between the viewports to take measurements. For cultures where the sedimentation rate was high, we mixed the sample every 30 seconds with a Pasteur pipette through the lid of the flask.

### 2.3.1 Calibration of plankton body size estimates

We compared body size estimates for small (5p0x) and large (0p5x) plankton organisms by the DSPC, featuring automated object detection and features extraction, to body size measurements done with images taken by inverted microscopy, using manual object detection and feature extraction using ImageJ (https://imagej.nih.gov/ij/). Having information on only two dimensions, size was evaluated as the area of objects rather than volume. For the comparison, only ROIs by DSPC that were comparable to manual microscopy (i.e. not cropped and imaged from the frontal plane) were selected and used for feature extraction (i.e. object area). Body size was estimated from DSPC images using SPC convert scripts (see Section 2.1.2). Taxon specific mean body size was calculated by taking an average area in pixels over all saved objects, both for DSPC and microscopy.

**5p0x magnification.** We used the following phytoplankton cultures: *Aphanizomenon flos aquae, Cyclotella meneghiniana, Scenedesmus acuminatus, Tetraedron minimum, Eudorina unicocca, Lagerheimia hindakii, Staurastrum punctulatum, Cosmarium botrytis, Oocystis solitaria, Planktothrix rubescens, Cystodinium sp, Volvox aureus, Peridinium sp.* All cultures were grown in WC medium (Guillard, 1975) under standard conditions (20°C and 10:14 light:dark cycle) for at least 2 weeks. Cultures of two small zooplankton were also imaged at 5p0x magnification, the protist *Euglena sp.* and the rotifer *Cephalodella sp.* These were grown under standard conditions and with a protozoan pellet medium (provided by CarolinaTM, Biological Supply Company, Burlington NC, USA). A volume of 2-10 mL (depending on initial culture densities) was inoculated into a 25 cm$^2$ tissue culture flask (vented cap, sterile, VWR 743-2311) and topped with tap water to a final volume of 60 mL.

**0p5x magnification.** We used the two large protists *Blepharisma sp. Spirostomum sp.,* which *w*ere grown under the same conditions as the protists used for 5p0x calibration. We also used living samples of *Copepods sp.*, *Daphnia sp.*, *Keratella quadrata*, *Keratella cochlearis* and *Kellicotia sp*. taken from Lake Greifensee (Switzerland).

**Traditional microscopy.** An aliquot of the imaged sample was taken and photographed with a light inverted microscope (Leica DMi8 magnification x160 or x320, depending on the size of the plankton species). Using ImageJ, the maximum and minimum axes length (perpendicular) were manually measured for 50 to 200 individuals, and then used to calculate the object area by applying an ellipses formula.

We compared the average number of objects detected per frame (i.e. per second) in the DSPC to density measurements obtained by inverted microscopy for small (5p0x) and stereomicroscopy for large (0p5x) plankton organisms, in serial dilutions. This DSPC metric is not sensitive to frame rate and we tested its comparability to volumetric measurements.

**5p0x magnification.** The following cultures were used: *Scenedesmus acuminatus, Oocystis solitaria, Tetraedron minimum, Botryococcus braunii , Euglena sp.* and *Cephalodella sp.* For this experiment, we used culturing flasks (as in the previous experiment). An initial sample of 0.5 to 10 mL of the culture was taken and topped to 60 mL with tap water. After imaging each culture for 5 minutes, 30 mL of the samples were replaced with tap water to prepare the next dilution step (1:2 dilution), followed by imaging with the DSPC. This was repeated for five dilution steps for *Scenedesmus acuminatus, Oocystis solitaria, Tetraedron minimum, Botryococcus braunii* and six dilution steps for *Euglena sp.* and *Cephalodella sp.* Images were then sorted out manually, whereas cropped or wrong images were not considered. The initial culture density was estimated from a subsample counted manually under an inverted microscope. **0p5x magnification.** Cultures of three Daphnia species, of different size classes*: D. magna, D. longispina, D. cucullata* were used. They were diluted (1:2) in six decreasing levels from 200 to 7 individuals (counted manually), and imaged in a culturing flask (total volume of 60 mL) for 5 minutes each. The images were manually sorted: first, all pictures that were considered as *Daphnia*, including cropped or/and not in focus were used for density estimation. Second, for the size estimation, only pictures of *Daphnia* in focus and not cropped were considered.

## 2.4 Field deployment and comparison with traditional monitoring

### 2.4.1 Automated deployment and monitoring station

The dual-magnification DSPC was installed at our automated monitoring station located in Greifensee, a peri-alpine, eutrophic lake in Switzerland (47.35 °N, 8.68 °E). Specific details about the automated monitoring platform can be found in (Pomati et al., 2011). The station features a multiparametric probe Ocean7 and an automated profiler (https://www.idronaut.it/), a meteorological station (WS700-UMB from OTT Hydrometrie /

LUFFT), a local computer and data network by 4G modem allowing data transmission. All data are streamlined and published in real-time on our website [www.aquascope.ch](www.aquascope.ch).

### 2.4.2 DSPC settings and data

The DSPC is permanently installed at a depth of 3 m, monitoring the plankton community hourly, for 10 minutes at the start of every hour, at an imaging rate of 1 frame / second. **Phytoplankton:** for comparisons with traditional monitoring, we used only the 5p0x magnification and selected the images of the DSPC that were taken in the same hour as a microscopy sample. This was mostly between 10:00-12:00 in the morning. The whole 10 minutes of running time and associated images were used. The images were then classified taxonomically to the lowest possible classification level. **Zooplankton:** We used only the 0p5x magnification data and aggregated discrete hourly samples over an entire day to allow for comparisons with traditional twin-net tows. Specifically, we took the mean of a 24 h running time, 10 minutes every hour. This was to increase DSPC sample size, since net-tows concentrate hundreds of liters of water. The images of 0p5x magnification were subset to every 6 second in order to exclude multiple imaging of the same object, since some individuals were observed to remain in the focal plane of the camera for some seconds, particularly when the water was calm. The images were then classified at the lowest possible taxonomic level using the trained CNN described in Section 2.2.

### 2.4.3 Traditional sampling and microscopy

Water samples were taken every week for 12 consecutive months starting in April 2019, with the exception of winter months (November - early March 2020) when we sampled fortnightly. All samples were taken between 10:00 and 12:00 in the morning. **Phytoplankton:** Water samples were taken near the DSPC, at 3 m depth, with a 5 L Niskin bottle. An aliquot of 50 mL was fixed with Lugol's iodine solution, and then identified and counted in the laboratory under an inverted microscope (Utermöhl method). **Zooplankton:** Integrated water column samples, starting from January 2020 onwards, were taken with a 95 μm twin net, from 17 m to surface. Samples were concentrated with a 95 μm mesh and fixed in 50 mL of 100% ethanol. Identification and counting was done under a stereomicroscope.

## 2.5 Data analysis

All data analyses and plotting were performed in the R programming environment ([https://www.r-project.org/](https://www.r-project.org/)). All the laboratory and monitoring data from this study will be made available from a public repository before publication (link to the repository will be reported here).

### 2.5.1 Laboratory calibration

**Body size:** We calculated the size of objects as 2 dimensional areas of the object mask, as automatically provided by our image processing scripts for DSPC, or manually per each taxon in microscopy. We used a least-square linear regression to correlate microscopy measurements (the benchmark) and measures taken by the DSPC, in $Log_{10}$ space. Size-density curves were done by using geom_density from the ggplot2 package on untransformed areas (geom_density, ggplot2 3.2.1). **Density:** Measurements were $Log_{10}$ transformed, with zeros set to an arbitrary low value (0.1) before transformation. When dilution series were run multiple times, we took the mean between technical repeats. Density estimates by microscopy (the benchmark) and DSPC were then compared using linear regression for each taxon. Similarly, an overall relationship between microscopy and DSPC was calculated with linear regression in $Log_{10}$ space for each magnification.

### 2.5.2 Field application

**Seasonal patterns:** We estimated the mean size of the phytoplankton community (as area of ROIs detected by the 5p0x magnification), and its densities, using the DSPC data from 2018. As a comparison to DSPC density estimates, we used the Chlorophyll-a (Chl-a) concentration recorded by the CTD-probe fluorimeter (https://chelsea.co.uk/products/trilux/), using averages of Chl-a measurements from 2.9 to 3.1 m from each profile. We classified objects at 0p5x magnification, including zooplankton taxa and several large colonial phytoplankton, with the trained CNN using images sampled from the full dataset every 6 seconds, again to avoid multiple images of the same object. Counts were then aggregated into coarser taxonomic levels (copepods, daphnids, rotifers, predators and nauplia) for comparison with DSPC data. **Day-night patterns:** To study phytoplankton daily division (cells grow during the day and divide during the night) and consequent change in community average size, and zooplankton community day-night migrations, we used phyto- and zooplankton body size (area) and abundance (ROI / second) data over 24 h across 34 non-consecutive days extracted with a weekly interval from the data collected in 2020. Again, 0p5x data were down sampled to one frame every 6 seconds, then ROIs were classified with the CNN model, and finally aggregated into coarse taxonomic groups. **Community diversity:** Community diversity metrics were estimated using ROI counts performed manually (classification and counting of images) for the 5p0x magnification, and automatically (CNN) for the 0p5x magnification. For both comparisons, the number of data-points used amounted to the weekly samples taken for traditional plankton monitoring methods (n = 44 for phytoplankton and 30 for zooplankton). Richness, Shannon diversity and Pilou's evenness were calculated using functions of the R package vegan (2.5-6) and linear density data. Relationships between traditional monitoring and DSPC in terms of plankton abundance and diversity were estimated using least-square linear regression.

# 3. Results

## 3.1 Laboratory calibration

To assess the DSPC sensitivity in automatically detecting objects, how such detection scales with plankton densities, and its accuracy in extracting object features such as plankton body size (i.e. the area of the object in two dimensions), we performed laboratory experiments to compare DSPC results with traditional (manual) microscopy methods. We used phytoplankton cultures, protists cultures, daphnia cultures and fresh field samples for rotifers, copepods and other zooplankton species. We selected species to represent a wide range of sizes, from *Tetraedron minimum* (0.00018 mm$^2$) to *Daphnia magna* (3.2 mm$^2$).

Overall, we found a strong linear relationship in Log-Log space (Log$_{10}$, R$^2$ = 0.959) between inverted microscopy (area calculation based on ImageJ) *versus* automated imaging and ROI area extraction using the DSPC (**Fig. 2**). The relationship indicates that size estimates tended to be larger in the DSPC. For both magnifications, we observed a stronger deviation from the linear relationship at the smaller size end. **Fig. 2A** shows how the lower resolution for smaller taxa in DSPC images makes small objects appear overpixeld and blurry, biasing the automated edge detection and therefore area estimation. The only taxon lying below the 1:1 line was *Volvox aureus*, a very large phytoplankton colony for which we detected many partial images. Individual size distributions for each taxon confirmed an overestimation of size (as ROI area) at the lower size end. In both magnifications, density distributions of size estimates overlapped better for larger taxa, i.e. *Daphnia*, *Cephalodella*, *Oocystis*, compared to smaller taxa, i.e. *Keratella quadrata*, *Scenedesmus acuminatus*, *Tetraedron minimum* (**Supplemental Fig. S1**).

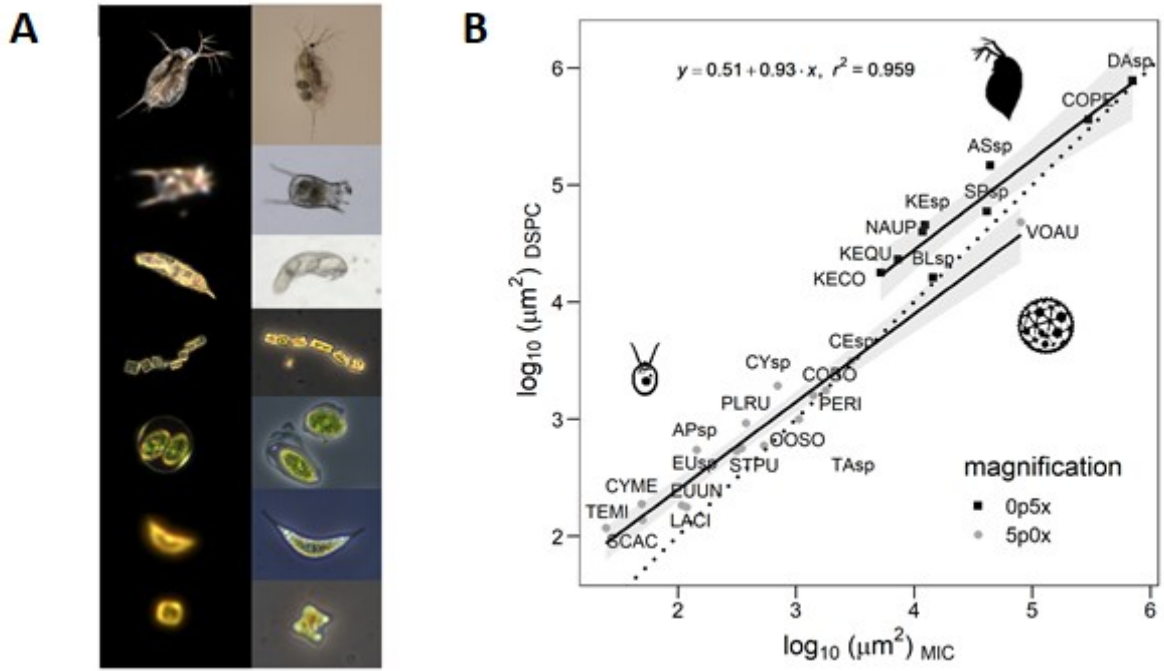

**Fig. 2**. **Relationship between body size estimates by inverted microscopy and the DSPC**. [A] Selected pictures of species by DSPC (left) and inverted microscopy (right), sorted based on size from the biggest taxon (on top) to the smallest (bottom). [B] Mean body size (ROI area) between inverted microscopy (MIC) and the plankton camera (DSPC). Dotted line = 1/1 relationship, solid lines = linear regression estimates ($p < 0.05$) and 95% confidence intervals.

For all taxa, density estimations by microscopy scaled significantly and accurately with ROI/s detected by the DSPC (**Fig. 3** - $R^2$ = 0.89 and $R^2$ = 0.73 for phytoplankton and zooplankton, respectively). We compared the number of objects detected per second (i.e. per frame) in the DSPC to density measurements obtained by inverted microscopy for small (5p0x) and stereomicroscopy for large (0p5x) plankton organisms, in serial dilutions. We performed serial dilutions of zoo- and phytoplankton taxa ranging of different sizes (118 um$^2$ - 3.2 mm$^2$). The overall scaling was more consistent across taxa for 5p0x magnification than for 0p5x. However, the strength of the relationship did not depend on size.

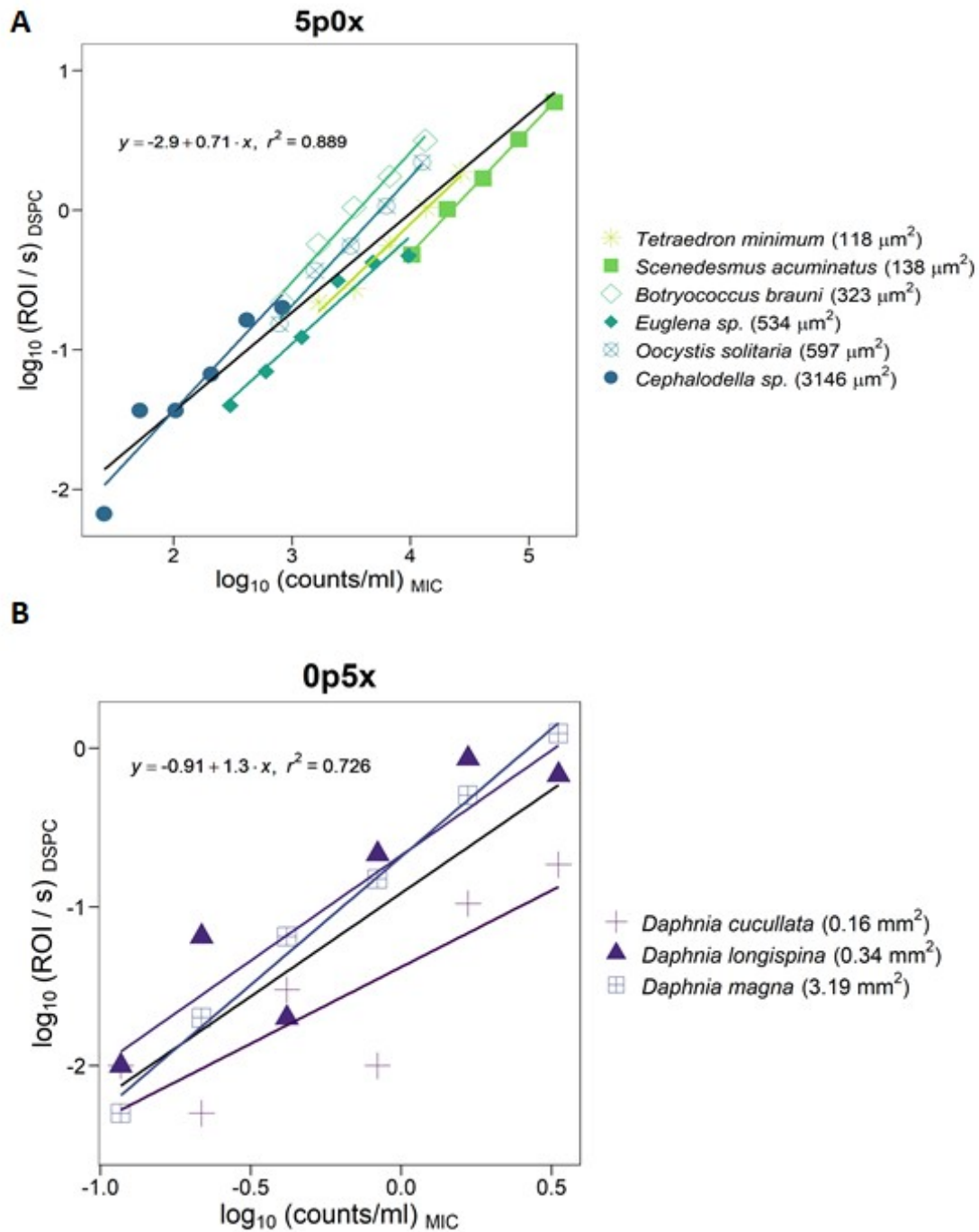

**Fig. 3. Calibration of abundance estimates by DSPC relative to microscopic counts.** [A] Phytoplankton and small protists ordered according to body size (area, measured by DSPC). [B] Three different species of *Daphnia* ordered according to body size (area, by DSPC). Black lines show the overall relationship between $Log_{10}$ transformed abundance measures by the DSPC (ROI / s) to density estimates based on inverted microscopy (counts / mL). Colored lines and dots represent the regression lines for the single taxa. All relationships are significant to $p < 0.05$.

## 3.2 Field deployment and biodiversity dynamics

To explore the potential of the new *in situ* imaging approach for automated monitoring of plankton community dynamics, we installed a dual-magnification DSPC permanently, at the stationary depth of 3 m, at the Eawag platform in the middle of Lake Greifensee (Switzerland) (www.aquascope.ch). The instrument has been operating autonomously, with rare interruptions of a few days a year for maintenance, since April 2018. During this time, we have acquired imaging data to describe the dynamics of the planktonic food web at high temporal frequency, and regularly taken samples with traditional monitoring approaches to compare biodiversity measurements by the DSPC to standard sampling and analysis methods. Patterns collected between 2018-2020 highlight seasonal successions and bloom dynamics, circadian rhythms in abundance and size. Comparisons with traditional methods reveal advantages and limitations of this new approach.

### 3.2.1 DSPC characterization of plankton diversity

In over two years of data collection in Lake Greifensee we could identify 28 plankton taxa with the 0p5x magnification (mostly zooplankton) and 80 taxa with 5p0x magnification (mostly phytoplankton). Some taxa were present in both magnifications, i.e. ciliates, rotifers and large phytoplankton colonies. **Fig. 4A** displays a collage of images representing the diversity of plankton communities as was observed with the dual-magnification DSPC, which encompassed primary producers, mixotrophs, herbivore grazers and carnivore zooplankton. Qualitatively, the high resolution and dark field background of images made taxon identification possible and facilitated image processing and training of a deep-learning classifier for zooplankton (see further Sections), with both morphology and color playing an important role.

The DSPC imaging approach represented a non-invasive and unbiased method to study plankton in the field, since taxa and their microenvironment were monitored without disturbing natural behavior, heterospecific aggregates and colonies structures (**Fig. 4B-D**). Some potentially toxic algae, such as the cyanobacteria *Mycrocystis* or *Planktothrix*, were successfully imaged and clearly recognized (**Fig. 4B**). Apart from colonies and natural aggregates (such as lake snow - **Fig. 4C**), we observed direct interactions between, for example, hosts and parasites (**Fig. 4D**) or predators and prey (**Fig. 4F**). Some of these images might be the first ever to capture direct individual trophic interactions in freshwater microbes from the field, such as ciliates and rotifers preying on phytoplankton (**Fig. 4F**). Colored images enabled us to distinguish between living and dead cells, and to observe plankton division and reproduction, i.e. by imaging dividing cells or individuals carrying eggs (**Fig. 4E**). For some large zooplankton like *Daphnia*, it is possible to count eggs from images.

The images also allowed us to identify and count different life stages of zooplankton, as is the case for juvenile copepods and nauplii.

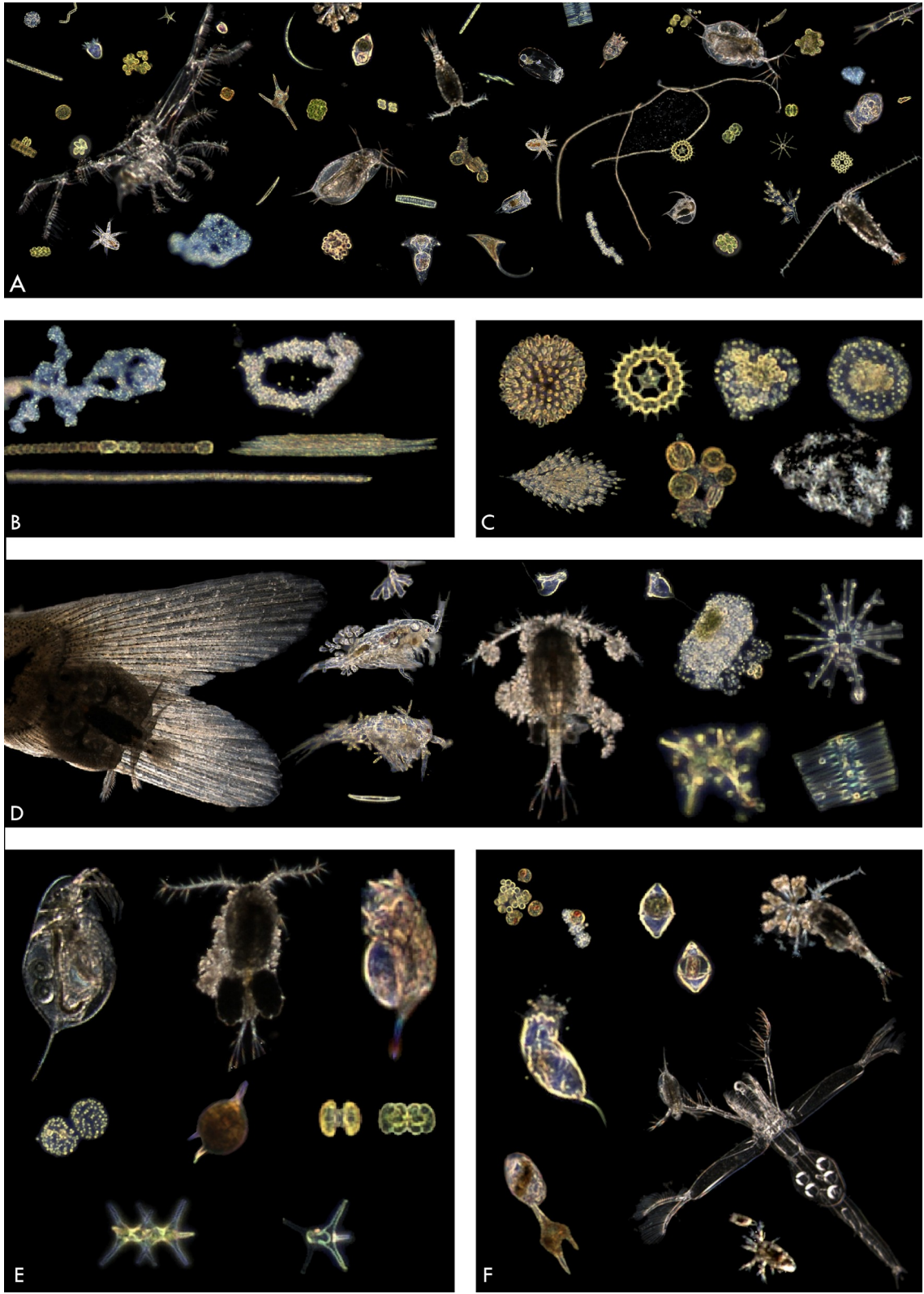

**Fig. 4. DSPC images of plankton taxa and natural taxa aggregates from Lake Greifensee (pictures are not to scale)**. [A] Collage of different images describing the large range of taxa and forms that can be imaged with the dual-magnification DSPC. [B] Potentially toxic cyanobacteria, left to right, top to bottom: two colonies of *Microcystis sp., Dolichospermum sp., Aphanizomenon flos-aquae and Planktothrix rubescens*. [C] Colonial algae - left to right, top to bottom: : *Uroglena sp., Pediastrum sp, two colonies of Coelosphaerium,* (bottom row) *Dinobryon sp*., natural aggregates of centric diatoms and aggregates of *Asterionella sp.* [D] Hosts and parasites (or epibionts) - left to right, top to bottom: parasite on a fish fin, *Epistylis sp.*on copepods, chytrid fungi on *Asterionella formosa* and *Fragilaria crotonensis* colonies, and unknown epibionts on *Staurastrum sp*. [E] Reproduction and life stages from left to right, top to bottom: *Daphnia sp., Cyclops sp. and Keratella cochlearis* (with eggs), *Coelosphaerium sp*., a cyst of *Ceratium hirundinella*, two dividing stages of C*osmarium sp., two Staurastrum sp*. - a dividing one and a dying individual. [F] Individual trophic interactions - from left to right, top to bottom: dinoflagellates eating green algae, two *Peridinium sp.* after eating centric diatoms, *Cyclops sp*. Eating a rotifer colony, a rotifer eating algal aggregate, *Leptodora kindtii* hunting for a Daphnia, *Ceratium hirundinella* interacting with a ciliate and a copepod nauplius interacting with an unknown organism.

## 3.2.2 Comparison with traditional sampling and microscopy

We compared phytoplankton and zooplankton biodiversity metrics estimated by the DSPC at 3 m to those obtained by traditional monitoring methods. The taxonomic identification in DSPC images was variable depending on the taxon and on whether the characters that are used for classification were distinguishable from the images (**Supplemental Tab. 5**). For both magnifications, taxonomic classification was harder at the lower end of the detection spectrum of the cameras, but was on average possible at the genus level. Phytoplankton richness estimates by the two approaches, however, agreed well overall, considering the differences in both sampling and analysis methods, with DSPC data systematically underestimating richness compared to traditional methods (**Fig. 5A**). Temporal trends in phytoplankton richness highlighted this offset, which however varied over time likely depending on phytoplankton community composition: data showed a better agreement between the two approaches in winter, when fewer and easily detectable taxa, such as Cryptomonas sp., were present (**Fig. 5B**). Zooplankton communities are in general less rich than phytoplankton, and both approaches detected low richness levels in our samples (max. 11 species). We did not find a significant correlation between zooplankton richness estimates by DSPC and traditional methods, and the relationship appeared even negative (**Fig. 5C**). In the case of zooplankton richness estimation, the DSPC, which averaged data over a 24 h period, detected higher richness than the traditional method based on a daytime net-tow and microscopy.

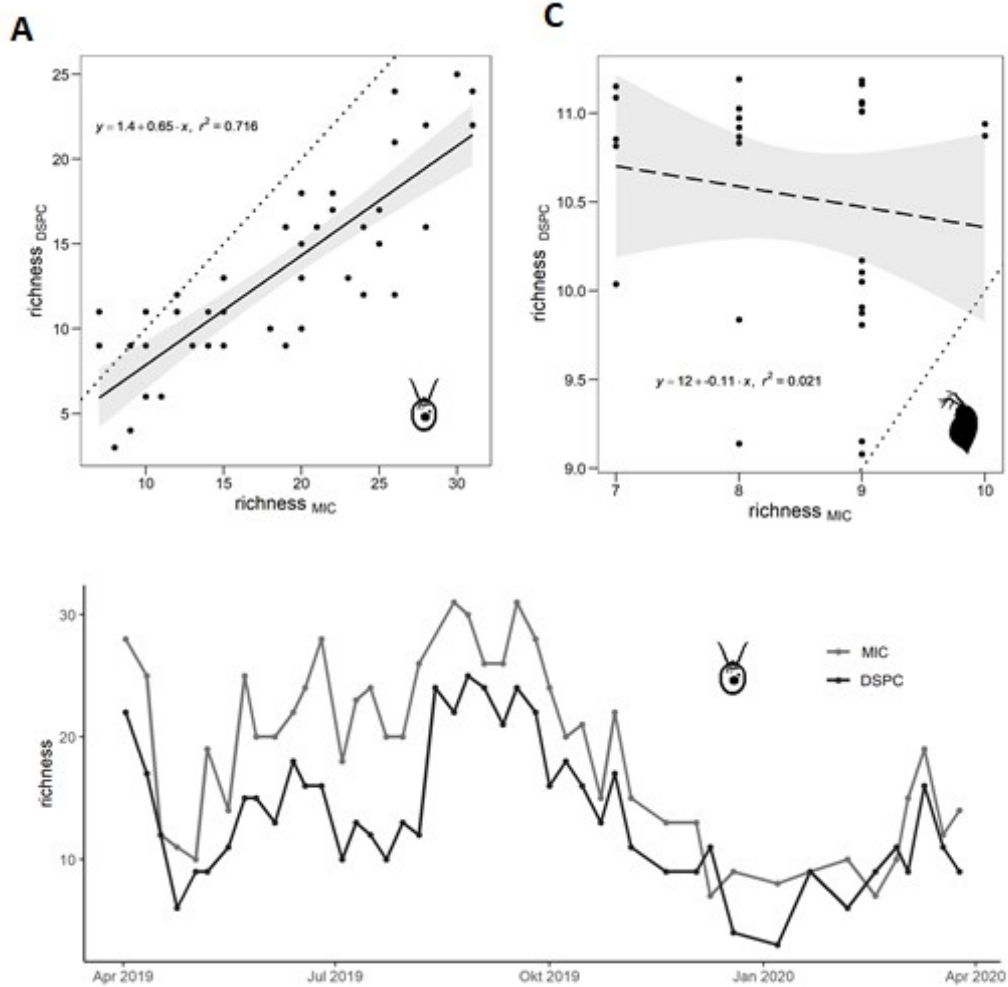

**Fig. 5. Comparison of species richness from DSPC relative to traditional methods.** [A] Phytoplankton richness (DSPC from 5p0x magnification - manual counting of images). [B] Zooplankton richness (DSPC from 0p5x magnification - manual counting of images). [C] Phytoplankton richness by DSPC compared to weekly traditional plankton monitoring data during 2019-2020. Dotted line = 1/1 relationship; solid lines = linear regression estimates (p < 0.05) and 95% confidence intervals; dashed lines = not significant (p > 0.05).

Total phytoplankton and zooplankton community densities estimated by the DSPC agreed well with data from traditional sampling methods (**Fig. 6A-B**). The phytoplankton relationship was statistically significant and robust: ROI/s generated by the DSPC correlated strongly also with Chl-a concentrations measured *in situ* next to the instrument (**Fig. 5** and **Supplemental Fig. S2**). For zooplankton, the relationship between density estimates by daytime net-tows and DSPC 24 h averages was statistically significant, but poor (**Fig. 6B**). The relationship between phytoplankton taxa relative abundances (i.e. evenness) by DSPC and traditional methods was statistically significant but poor, with data however scattered around the 1/1 line (**Fig. 6C**). Zooplankton taxa evenness, instead, showed a significant and relatively strong relationship between the two approaches (**Fig. 6D**).

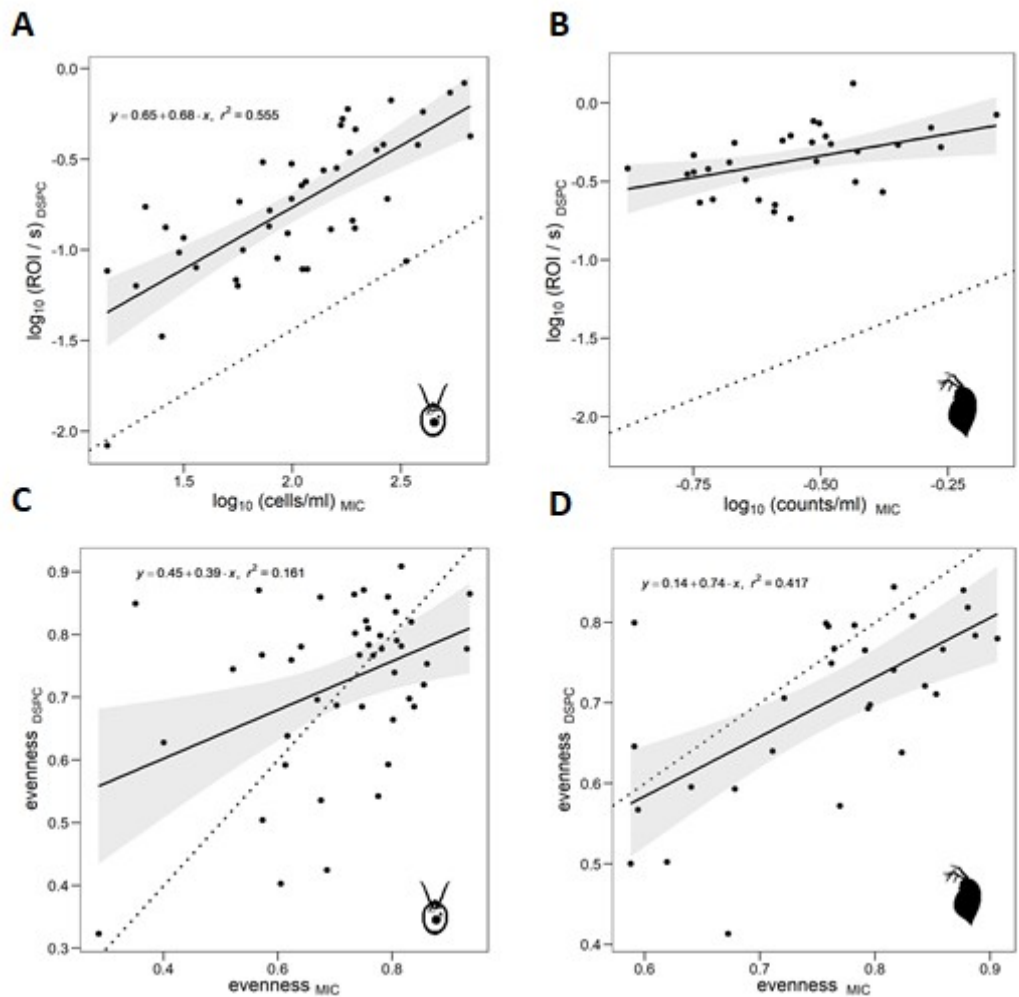

**Fig. 6. Comparison of total and relative abundance estimates by DSPC to traditional methods.**
[A] Total phytoplankton community density. [B] Total zooplankton community density. [C] Phytoplankton relative taxa abundances (Pielou's evenness). [D] Zooplankton relative taxa abundances (Pielou's evenness). Density data were $Log_{10}$ transformed for A and B. Pielou's evenness in C and D was calculated using linear density data. DSPC data: phytoplankton from the 5p0x magnification and zooplankton from 0p5x magnification (considering only crustaceans) - images were manually counted. Dotted lines = 1/1 relationships in C and D, and calibration relationships from laboratory serial dilutions in A and B (Fig. 3), solid lines = linear regression estimates (p < 0.05) and 95% confidence intervals.

### 3.2.3 Seasonal patterns

The DSPC data in 2018 depicted the largest algal bloom in Lake Greifensee since the early 1990s in terms of phytoplankton densities (see Chl-a levels in **Fig. 7A**). ROI/s collected by the DSPC for the 5p0x magnification correlated highly with the Chl-a measured by the CTD fluorimeter at 3 m depth ($R^2$ = 0.94, *p* < 0.001; **Supplemental Fig. S2**). Both Chl-a concentration and ROI/s by the DSPC (**Fig. 7**) showed a relatively small increase in phytoplankton at the end of July, and then a large accumulation in the middle of September, which was characterized by a relatively smaller mean phytoplankton community size (ROI area) compared to non-bloom periods (**Fig. 7A**). The late summer bloom was dominated by the small single-cell taxon *Oocystis sp.*, as identified by DSPC images, while pre-bloom communities were more heterogeneous (**Supplemental Fig. S3**). Patterns of co-occurring zooplankton taxa and their abundance (ROI/s) showed that relatively high density of herbivore grazers (particularly daphnids) in the pre-bloom phase (middle of August) was followed by a decrease in their abundance after the DSPC detected the presence of carnivore species (particularly *Leptodora*) at the onset of the phytoplankton bloom (**Fig. 7B**). This release from top-down grazing pressure might have been involved in the emergence of the bloom, which was dominated by a small edible taxon. At the end of September, the algal bloom declined as we observed an increase in small fast-growing herbivores (i.e. rotifers), which might have been involved in bloom termination (**Fig. 7B**).

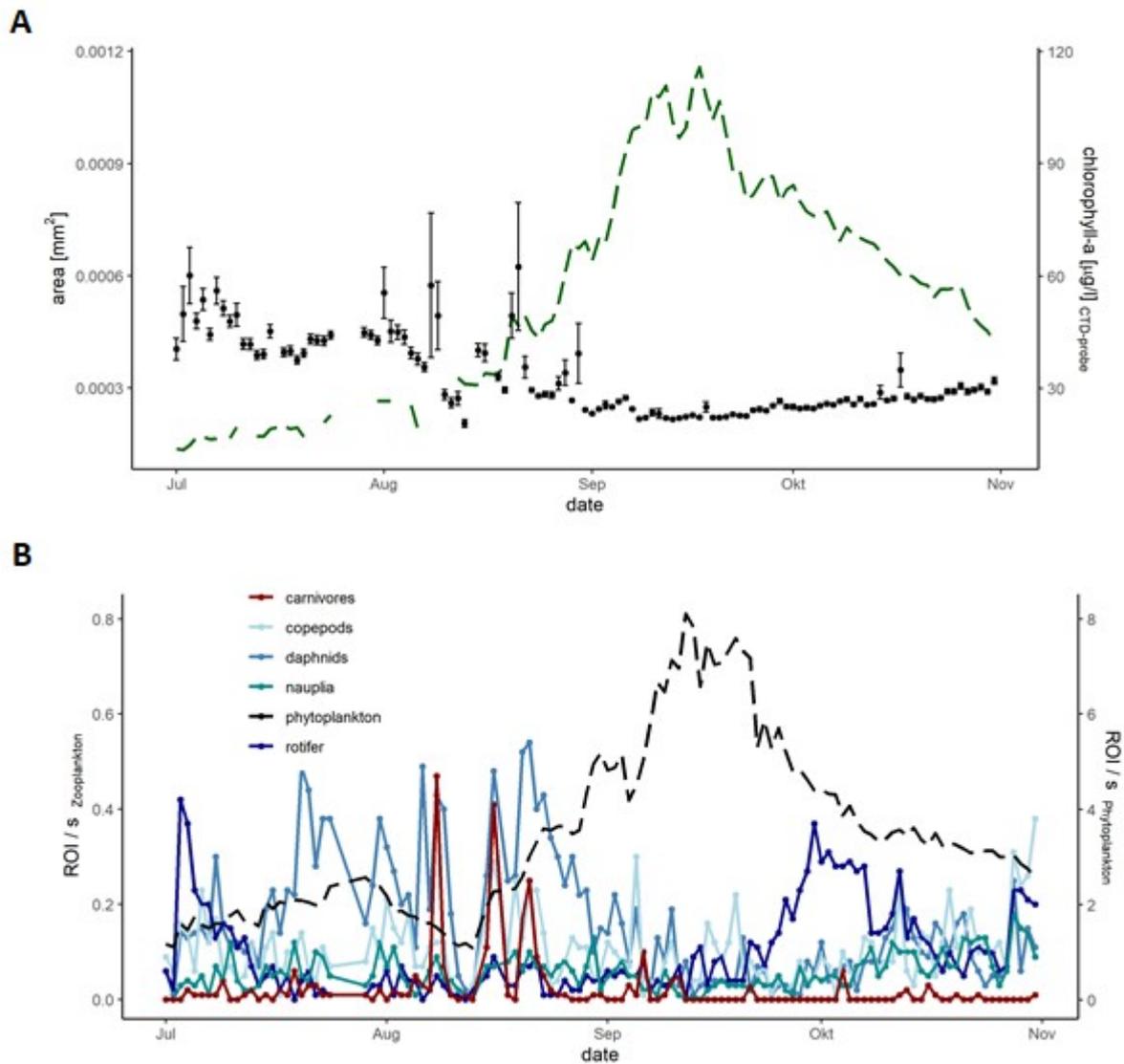

**Fig. 7. Patterns from DSPC monitoring data from Lake Greifensee (2018).** [A] Chl-a data from the CTD and fluorimeter (dashed green line), and phytoplankton community size by DSPC (points represent the mean ROI size in the 5p0x magnification +/- standard error of the mean). [B] Patterns in plankton density (ROI/s) dynamics from the DSPC: colored lines represent different zooplankton taxa (based on CNN automated classification, from 0p5x magnification, see **Supplemental Table S3**), dashed black line represents phytoplankton (from the 5p0x magnification). Data represent one DSPC measurement per day (10 min running time, subset to 6 seconds for zooplankton) taken during the night (01:00).

### 3.2.4 Circadian patterns

Time series generated by the DSPC, which was set for 10 min of measurements every hour, 24/7, allowed study of circadian size dynamics and of zooplankton day/night migratory behavior (**Fig. 8**). Phytoplankton mean community size showed to be slightly larger during the day compared to the night (when both grazing and cell division mostly occur) (**Fig. 8A**). Zooplankton mean community size was larger, instead, during the night than during the day (**Fig. 8B**). This was likely due to different migration behaviors of large relative to small herbivore grazers (**Fig. 8C-F**). Copepods and daphnids were more abundant (higher ROI/s) during the night, and peaked between midnight and 3 am (**Fig. 8C-D**). During the day, daphnids were almost undetectable. Conversely, nauplia and rotifers (the most abundant group at 3 m) did not show any significant circadian patterns (**Fig. 8E-F**).

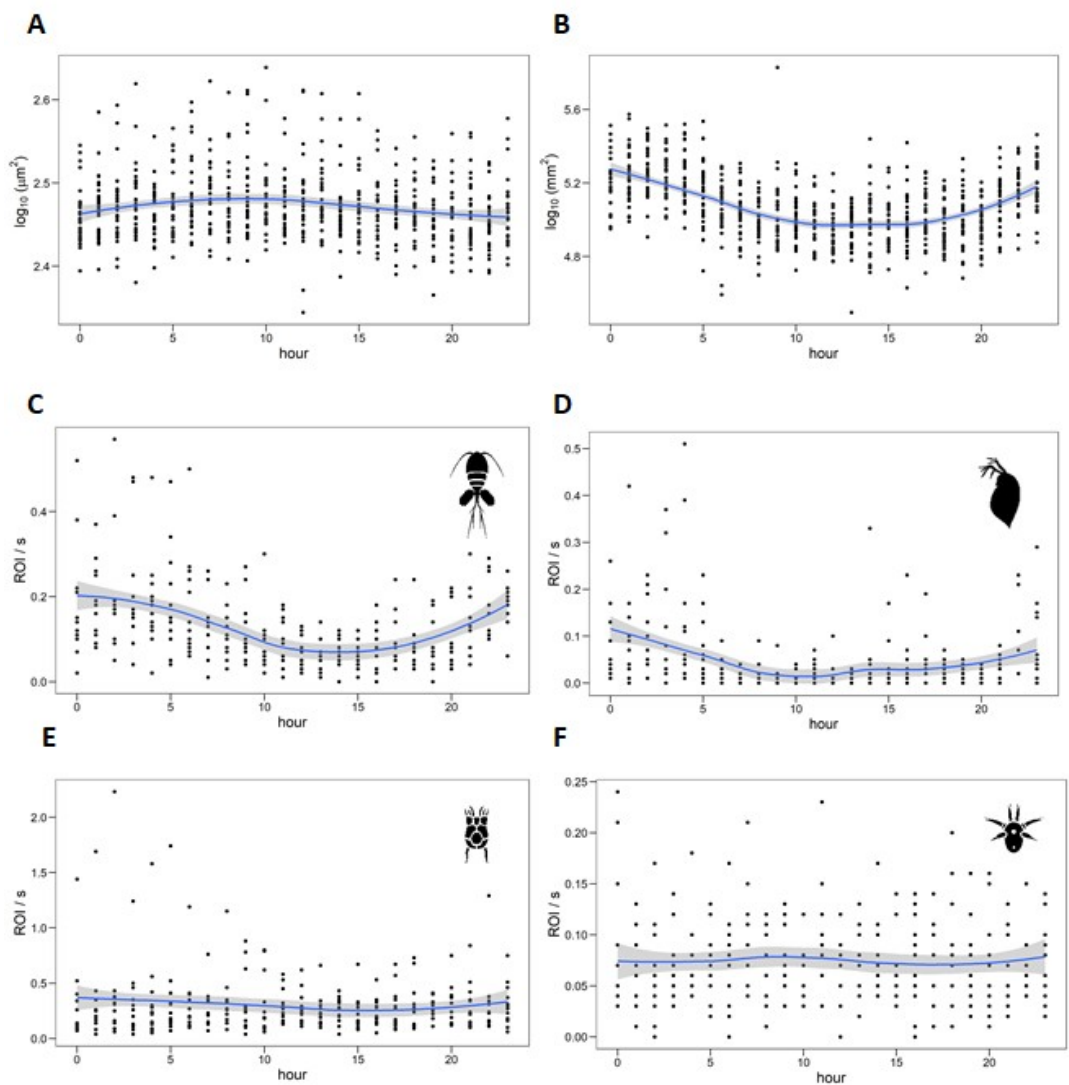

**Fig. 8. Circadian patterns in plankton size and abundance.** Change in phytoplankton community mean size over 24 hours (from 34 days with a weekly interval during 2019 and 2020) [A]. Zooplankton community mean  size change over 24 hours [B], and density (ROI/s) of copepods [C], daphnids [D], rotifers [E], and nauplia [F] (from 34 random days in 2020). Zooplankton counts based on CNN automated classification, from 0p5x magnification, see **Supplemental Table S3**.

# 4. Discussion

The dual magnification DSPC allowed imaging plankton individuals, colonies, and aggregates of taxa and other suspended solids (e.g. lake snow) without disrupting natural arrangements of interacting organisms, their micro-environment or their behavior. The instrument reliably delivered high-resolution images in real-time, and measurements of traits like size, shape and color. This included detection capability across a wide-reaching size range of the planktonic food web.  Detecting, with high resolution, particles from nanoplankton to mesoplankton, is a dynamic range never achieved before in other *in-situ* imaging systems (Lombard et al., 2019). Taxonomic identification was variable depending on detected taxon and the resolution of the image. In general, however, the colored images with dark background afforded high object definition and contrast, qualities that facilitate the identification of taxa and the annotation of ROIs for training of deep-learning models. Automated image classification was achieved for plankton with the 0p5x magnification in this study (**Supplemental methods**) and elsewhere (Campbell et al., 2020; Orenstein et al., 2020). The training of a phytoplankton classifier will be manageable with relatively low effort (Kenitz et al., 2020), however more tedious due to the larger number of taxa to annotate. An approach based on imaging has the potential to yield high frequency and standardized plankton monitoring data (Lombard et al., 2019) and, boosted by continuous (and open-source) advances in image processing and artificial intelligence, offers the prospect to automate and standardize also taxonomic classification of plankton (MacLeod et al., 2010).

Time series collected by the stationary DSPC in Lake Greifensee depicted seasonal succession patterns, bloom dynamics and circadian fluctuations with a temporal and spatial (i.e. microenvironment and morphology) resolution that would have been virtually impossible to collect based on traditional methods. In particular, the DSPC data included dynamics of all the planktonic groups that are important to understand and predict community dynamics (including algal blooms), but are often overlooked in routine monitoring (e.g. ciliates, chytrid fungi) (**Fig. 2**). In addition, for all the monitored plankton taxa, automated underwater imaging allows following the dynamics of the distribution of important traits such as size, shape, and pigmentation, which are extremely hard to measure manually on large numbers of individuals. Morphological characteristics of phytoplankton such as length, surface, coloniality, and pigmentation correlate with species functional properties, such as resource acquisition, growth rate, sinking rate, grazing resistance, and therefore with population abundances and responses to environmental gradients (Fontana et al., 2019; Kruk et al., 2010; Pomati et al., 2013).

In this study, DSPC imaging and feature extraction allowed calculation of bi-dimensional areas for ROIs, which related well to areas of plankton taxa based on traditional microscopy and manual measurements. Although measured in two dimensions, the body size estimates by one or two linear dimensions scale with biovolume estimates by traditional methods (Mittler et al., 2019). Body size was overestimated by the DSPC at the lower size end, where both regression lines for the 0p5x and 5p0x magnification deviated for smaller species from a perfect 1/1 relationship. This was due to a lack of resolution in pixels for smaller individuals, particularly when slightly out-of-focus, which can be corrected using the relationship derived by laboratory calibration (**Fig. 2**). The overestimation of size would be even greater when using estimates of 3-D biovolume (rather than 2-D area), so care should be given when calculating biovolume estimates.

Similarly, our measure of plankton densities (ROI / s, or alternatively ROI / frame), can be translated into a number of individuals per unit of volume (e.g. L) based on the laboratory calibration (**Fig. 3**). Note, however, that the initial densities in our laboratory experiments were much higher than those usually found in nature: for the goal of an *absolute quantitation* of plankton densities, we suggest a laboratory calibration targeted towards specific instrument settings and realistic densities. Measuring densities based on detected ROI / s would be suitable to estimate *relative abundances*. Other issues emerging from the comparison of DSPC data with traditional plankton sampling include: i) the difference between methods in terms of sampled volume, which is used to estimate diversity metrics; ii) no detection of phytoplankton < 10 µm, with loss of some small but common species; iii) the spatial heterogeneity and circadian behavior of zooplankton.

The DSPC imaged volume (**Supplemental Table S1**) is small compared to traditional monitoring samples, and variable depending on water optical conditions (Orenstein et al., 2020). The DSPC cannot easily change sampled volume to respond to changes in densities, unlike traditional counting methods. While we can apply a space for time substitution to increase imaging sampling effort, it is difficult to assess the actual differences in sample size between DSPC and traditional methods. This limitation, which applies to almost all automated *in-situ* methods, has been already acknowledged (Lombard et al., 2019). While automated imaging data allow the scaling of measures of diversity based on sample size (e.g. by temporally aggregating data and applying rarefaction and probability estimate of metrics) (Chase and Knight, 2013), this is generally impossible with traditional sampling methods. Those differences in scaling hamper our understanding of why imaging and traditional methods sometimes fail to converge to the same diversity estimates (**Fig. 5-6**).

The DSPC showed difficulties in detecting small species for the 5p0x magnification, and in quantifying densities of large rare species, including carnivorous zooplankton, for the 0p5x magnification. The lack of taxonomic resolution for small phytoplankton taxa (in 5p0x) is a limitation of the optical system itself, solvable by an extra (higher) magnification. The correlations between DSPC and microscopy estimates for phytoplankton richness and total abundance were however robust (**Figs. 5A** and **6A**), with the limitations in taxonomic resolution becoming more evident in the comparison of relative abundance estimates (**Fig. 6C**). The lack of reliable quantitation of large zooplankton taxa in the 0p5x magnification, instead, resides in the relationship between spatial density and body size of organisms (White et al., 2007). Imaging with the DSPC for a longer time, particularly at night when these organisms are more active, might ameliorate this problem by increasing the sampling effort. Due to a relatively small focal volume, however, large individuals tend to be imaged and cropped multiple times. We suggest that large carnivorous zooplankton can only be studied with presence/absence data or during periods when their densities reach very high levels.

Zooplankton species have patchy distributions and exhibit diel vertical and horizontal migrations, which could lead to systematic biases in measuring community and population properties, if these are based on non-replicated local samples, or only daytime samples (Doubek et al., 2020). We propose that the poor correlation between DSPC and traditional sampling for zooplankton richness and abundances was caused by the heterogeneous distribution of taxa over space and time, evident also by high levels of variation among replicated samples, particularly for traditional methods (**Fig. S7**). In this perspective, DSPC data (which have been integrated over 24 h and include both day and night samples) may be more trustworthy than the traditional daytime net-tows, which have been previously suggested not to be an appropriate approach for monitoring (Doubek et al., 2020).

Availability of high resolution real-time plankton data represent a potential breakthrough in both applied and fundamental plankton ecology and may be more important for water quality monitoring than a fine taxonomic resolution. High-frequency data have shown to provide exciting prospects for modelling, understanding and predicting plankton dynamics and ecosystems processes (Fontana et al., 2018; Thomas et al., 2018). The seasonal patterns depicted in **Fig. 7** highlight the rich description of the ecological succession of taxa that DSPC data can offer for studying plankton communities. Natural processes fostering high plankton biodiversity on one hand, and leading to the emergence of harmful algal blooms on the other hand, represent unresolved questions in aquatic ecology (Burford et al., 2020; Fox et al., 2010; Huisman et al., 2018; Li and Chesson, 2016). The DSPC was able to detect (and count) all the most common and potentially toxic cyanobacterial taxa forming blooms in lakes, including their large colonies (**Fig. 2**). If supported by an automated deep-learning classifier and paired with meteorological and CTD data (and/or a calibrated lake process models), the DSPC could be instrumental to develop forecasting models for harmful cyanobacterial blooms, and the stability and biodiversity of natural plankton communities (Pomati et al., 2011; Thomas et al., 2018).

Time series generated by instruments like the DSPC can be combined with novel data analysis approaches, including equation free modelling and machine-learning, to reverse engineer community ecology based on observational information and to develop effective forecasting tools (Deyle et al., 2016; Martin et al., 2018; Sugihara et al., 2012; Thomas et al., 2018). The combination of high-frequency data, broad taxonomic spectrum and artificial intelligence (for data assimilation and modelling), presents unparalleled potential for capturing the processes acting in natural ecosystems. This approach offers great opportunities to advance our understanding of the mechanisms controlling the dynamics of the complex networks of interacting plankton species by suggesting novel and realistic hypotheses to be tested experimentally and through theory development.

# 5. Conclusions

The approach presented in this article, based on dual-magnification underwater microscopy, allowed us to obtain automated plankton data at the food web level with high temporal frequency and adequate image resolution. The plankton camera employed here was demonstrated to be very robust and reliable, showed low requirements in terms of maintenance and running costs and offered complete automation. We conclude from our comparison with traditional monitoring methods that the approach is mature as a research tool, and suitable for stable continuous deployment, with great potential for application in water quality monitoring.

Automated *in-situ* imaging data are available in real-time, and they are shareable, traceable, transparent, and independently verifiable. Biases in the data processing and analysis can be corrected by simply re-processing and re-analyzing archived raw data. On the other hand, new-generation monitoring tools like the DSPC present the challenge of managing and analyzing large datasets. This might prove particularly relevant if underwater imaging cameras become more affordable and, to increase sampling effort or account for spatial variability in water bodies (both vertical and horizontal), multiple instruments are employed locally and used in full-time operation (e.g. to gain power for density estimates of rare taxa). Recent publications suggest that this transition from relatively expensive tools to more affordable, and therefore scalable, instruments is already underway (Lertvilai, 2020). Streamlined and open source data management protocols, image processing software and automated image classification algorithms are becoming more and more popular, and the approach that we propose can be employed for both fundamental and applied ecology, allowing calibration of lake ecological models, and supporting water quality monitoring with real-time plankton data.

# 6. Acknowledgements

We thank O. Köster, M. Koss and L. De Ventura for the fruitful discussions on the advantages and limitations of the imaging approach. This research was funded by the Swiss Federal Office for the Environment (contract Nr Q392-1149) and the Swiss National Science Foundation (project 182124). F.P and P. I. also acknowledge the Eawag DF project Cyanoswiss (#5221.00492.012.04). F.P. and M.B.-J. acknowledge the Eawag DF project Big-Data Workflow (#5221.00492.999.01).

# 7. Contributions

E.M., T.K., M.R., C.E., P.I., S.D. and F.P. planned the study. T.K. and M.R. performed field sampling and taxonomic classifications. C.E. installed the instrument in Lake Greifensee and ensured constant functioning. P.R. and J.S.J. designed and built the instrument. M.B.-J., T.L. and T.H. developed and trained the CNN. N.S. and S.D. created a script for zooplankton appendix erosion. E.M., T.K., M.R., P.I., M.B.-J., S.D. and F.P. drafted the manuscript. All authors provided critical feedback and approved the final version of the manuscript.

# Appendix. Supplementary Materials and Methods.

**Table S1**. Specifications of the plankton camera. The instrument is derived from SPCP2 and SPC cameras, for the high magnification http://spc.ucsd.edu/cameras-2/spcp2-camera/, and for the low magnification http://spc.ucsd.edu/cameras-2/spc-camera/, respectively (Orenstein et al., 2020).

| Camera | **Matrix Vision mvBlueFox3 (3.45um pixel size)** |
|---|---|
| Sensor | **Sony IMX253** <br><br> [*Point Grey GS3 12 MP, (GS3-U3-120S6C-C*) – this is the standard for SPC & SPCP. Eawag Pcam was custom made and has other hardware] |
| Objective Lens | 5p0x: **Olympus 5x Long Working Distance M Plan Semi-Apochromat (LMPLFLN5xBD)** <br><br> 0p5x: **Opto Engineering High resolution telecentric lens for 1″ detectors, magnification 0.508x, C-mount (TC2MHR024-C)** |
| Pixel Size (Object Space) | 5p0x: 0.62 um |
| | 0p5x: 6.2 um |
| Tube Lens | Thor Labs TTL200 |
| Computer | Nvidia Jetson TX1 |
| Operating System | Ubuntu 16.04, Linux For Tegra R24.2.1 |
| Power Requirements | 9-36VDC (AC Adapter provided) |
| Battery Pack and Controller | Inspired Energy 98 Whr (NH2054) with EB335 controller<br>About 3 hour runtime at full load |
| Depth Rating | Viewports: 50 m. Housings: 200 m<br>Original design tested at 200 m. |
| Power Input | 10-24 VDC; < 3A inrush @ 24V, Testing done at 24V. |
| Power Consumption | **0.2 W :** CTRL on, system off<br>**13 W :** CTRL on, system on, idle.<br>**25 W :** Full system load<br>Jetson consumes ~ 9 W, each camera consumes ~ 4 W, and UV LEDs consume about 7W. |
| Weight in Air | Approx 60 lbs (27.2kg) |
| Weight in Seawater | Approx 5 lbs buoyant (2.3kg) |
| Viewport Material | Acrylic due to limited resolution through sapphire for 5x magnification. |
| Housing Material | Acetal, Tubes (Natural) and Endcaps (Black) |
| End Plate Material | 6061-T6 Hard Black Anodize |
| Port Retaining Ring | 110 Copper |
| Bulkhead connectors | Subconn, brass |
| Anode material | Magnesium |
| External Hardware | Titanium Grade 2. Delrin spring lock washers |
| Strobes | Custom Darkfield Illuminator, CBT-140 Luminus LED for each |

| | magnification, roughly 4000k color temp. |
|---|---|
| Manufacturer Costs | ~ 80'000 $ (in 2018) |
| Video Performance | 10 FPS, raw 16 bit to disk, Approx 160 MB/s, per camera<br>Max 3800 x 2600 pixels in 16 bit mode |
| ROI performance | Up to 600 (small) ROIs per second to disk<br>Small ROIs are those will size roughly 100 x 100 pixels or less. |
| Detection range | 5p0x: ~10µm - 150 µm<br>0p5x: ~ 100 µm - 7 mm |
| Imaged volume | 5p0x: 0.2 - 10 µL / frame<br>0p5x: 4 - 200 µL / frame |

## CNN classifier for 0p5x images

The model we used for classifying zooplankton objects is a neural network with two convolutional layers, with kernel size (number of filters) of respectively 24 and 12 (64 and 32) and relu activations, followed by batch normalization and max pooling modules, and a dense layer with softmax activation[1]. The loss function was a categorical cross-entropy. We provide our code and trained model at:

https://github.com/mbaityje/plankifier/releases/tag/v1.1.1

The default code settings contain all the hyperparameter choices we made. The dataset we used consisted of 17909 images (**Table S2**), of which 80% were used for training, and 20% for testing. The training data and further benchmarks on this model, along with more sophisticated ones which reach higher performance, will be published separately[2]. The performances that we show in this section are based on further independent data.

**Table S2.** Classes and number of ROIs used for CNN training per class.

| Class | Number of images | Class | Number of images |
|---|---|---|---|
| dinobryon | 3321 | conochilus | 264 |
| nauplius | 1507 | trichocerca | 255 |
| maybe_cyano | 1364 | unknown | 245 |

---

1 Despite its simplicity, this model obtains satisfactory performances, and has the additional advantage of being very lightweight. This allows us to share it, already trained, in our github repository github.com/mbaityje/plankifier. The consequence is that practitioners can use it directly through the simple commands provided in the release documentation.

2 Kyathanahally, S., Merz, E., Hardemann, T., Reyes, M., Kozakiewicz, T., Isles, P., Pomati, F., Baity-Jesi, M., In preparation

| | | | |
|---|---|---|---|
| diaphanosoma | 1089 | aphanizomenon | 225 |
| asterionella | 1055 | fish | 222 |
| uroglena | 953 | keratella_cochlearis | 215 |
| cyclops | 866 | leptodora | 203 |
| ceratium | 814 | synchaeta | 142 |
| rotifers | 744 | dirt | 131 |
| daphnia | 721 | bosmina | 80 |
| asplanchna | 607 | polyarthra | 80 |
| eudiaptomus | 537 | unknown_plankton | 71 |
| kellikottia | 519 | daphnia_skins | 46 |
| paradileptus | 424 | copepod_skins | 33 |
| keratella_quadrata | 420 | hydra | 18 |
| filament | 405 | diatom_chain | 17 |
| fragilaria | 306 | chaoborus | 10 |

We rely on abstention to be able to tune the precision-recall tradeoff in a simple manner. This consists of labeling as unclassified all the images for which the classifier has a confidence lower than a threshold $\theta$. In **Fig. S7** we show the precision and recall of our model with $\theta$=0, on the classes that are relevant for our study. The average precision is 0.84, and the average recall is 0.8. If we set $\theta$=0.8, the average precision is 0.91, and the average recall is 0.67, which means that around 30% of the data is not being classified, but the data that gets classified has a very low error rate. The data shown in this paper uses $\theta$=0.8.

Since in this study we were interested in a higher-level taxonomic description of the observed organisms, we created some macro-classes, each comprising more than one of the classes described in **Table S3**. Obviously, the classifier performances with the macro-classes are at least as good as those of **Fig. S7**.

**Table S3.** Composition of the zooplankton functional groups used in the main text.

| Section in main text | Macro-class | Classes |
|---|---|---|
| **Fig. 5** | copepods | cyclops + eudiaptomus |
| | daphnids | diaphanosoma + daphnia + bosmina |
| | rotifer | kellikottia + keratella_quadrata + keratella_cochlearis + trichocerca + conochilus + asplanchna + rotifers |
| | predators | chaoborus + leptodora |
| **Fig. 6** | copepods | cyclops + eudiaptomus |
| | daphnids | daphnia + bosmina + diaphanosoma |
| | rotifers | kellikottia + keratella_quadrata + keratella_cochlearis + conochilus + asplanchna + rotifers + synchaeta + polyarthra + trichocerca |
| | nauplius | nauplius |

**Table S4**. Linear regression summary.

| figure | y-variable | x-variable | group | intercept | slope | r-sqr | p-value | significant |
|---|---|---|---|---|---|---|---|---|
| 1B | body size log10(um2)DSPC | body size log10(um2)MIC | 0p5x | 1.3529 | 0.7734 | 0.9038 | 0.00005202 | yes |
| 1B | body size log10(um2)DSPC | body size log10(um2)MIC | 5p0x | 0.8965 | 0.75061 | 0.9481 | 1.336E-10 | yes |
| 1B | body size log10(um2)DSPC | body size log10(um2)MIC | overall | 0.52642 | 0.92377 | 0.9572 | < 2.2e-16 | yes |
| 2A | density log10(ROI/s)DSPC | density log10(counts/ml)MIC | TEMI | -3.34127 | 0.8112 | 0.9732 | 0.001215 | yes |
| 2A | density log10(ROI/s)DSPC | density log10(counts/ml)MIC | SCAC | -3.87528 | 0.89204 | 0.9965 | 0.00005738 | yes |
| 2A | density log10(ROI/s)DSPC | density log10(counts/ml)MIC | BOBR | -3.30509 | 0.93043 | 0.9793 | 0.0008224 | yes |
| 2A | density log10(ROI/s)DSPC | density log10(counts/ml)MIC | Eusp | -3.2663 | 0.7702 | 0.9406 | 0.0008598 | yes |
| 2A | density log10(ROI/s)DSPC | density log10(counts/ml)MIC | OOSO | -3.4492 | 0.92219 | 0.9865 | 0.000435 | yes |
| 2A | density log10(ROI/s)DSPC | density log10(counts/ml)MIC | Cesp | -3.2581 | 0.9104 | 0.8886 | 0.003071 | yes |
| 2A | density log10(ROI/s)DSPC | density log10(counts/ml)MIC | overall | -2.86184 | 0.7107 | 0.889 | 7.40E-16 | yes |
| 2B | density log10(ROI/s)DSPC | density log10(counts/ml)MIC | DACu | -1.3782 | 0.9654 | 0.6431 | 0.03406 | yes |
| 2B | density | density | DALO | -0.6746 | 1.3255 | 0.7806 | 0.0123 | yes |
| | log10(ROI/s)DSPC | log10(counts/ml)MIC | | | | | | |
| 2B | density log10(ROI/s)DSPC | density log10(counts/ml)MIC | DAMA | -0.68311 | 1.61262 | 0.9894 | 0.0000269 | yes |
| 2B | density log10(ROI/s)DSPC | density log10(counts/ml)MIC | overall | -0.912 | 1.3012 | 0.726 | 7.153E-06 | yes |
| 5A | richnessDSPC | richnessMIC | | 1.4 | 0.65 | 0.716 | 4.81E-13 | yes |
| 5C | richnessDSPC | richnessMIC | | 11.5062 | -0.1149 | 0.021 | 0.444 | no |
| 6A | density log10(ROI/s)DSPC | density log10(counts/ml)MIC | | 0.65 | 0.68 | 0.555 | 1.254E-08 | yes |
| 6B | density log10(ROI/s)DSPC | density log10(counts/l)MIC | | -1.7346 | 0.5594 | 0.206 | 0.01181 | yes |
| 6C | evennessDSPC | evennessMIC | | 0.45 | 0.39 | 0.161 | 0.006888 | yes |
| 6D | evennessDSPC | evennessMIC | | 0.1403 | 0.7395 | 0.417 | 0.000115 | yes |
| S2 | density ROI/s DSPC | chlorophyll-a (ug/l)CTD-probe | | -0.23361 | 0.062248 | 0.9387 | < 2.2e-16 | yes |
| S4B | countsML | countsMA | rotifer | 0.42767 | 0.49976 | 0.7481 | < 2.2e-16 | yes |
| S4B | countsML | countsMA | daphnids | 1.32638 | 0.69017 | 0.8646 | < 2.2e-16 | yes |
| S4B | countsML | countsMA | copepods | 0.43969 | 0.55061 | 0.8221 | < 2.2e-16 | yes |
| S4B | countsML | countsMA | carnivores | 0.7658 | 0.32412 | 0.7389 | < 2.2e-16 | yes |
| S4B | countsML | countsMA | nauplia | 0.78086 | 0.65868 | 0.6551 | < 2.2e-16 | yes |

**Table S5**. **Phyto- and zooplankton taxa in Lake Greifensee.** 1 taxa was to some degree (genus or species) identified on images taken with the 5p0x magnification.

| Taxa MIC | Taxa DSPC | Seen in DSPC? | Comments |
|---|---|---|---|
| **Cyanobacteria** | | | |
| cyanobacteria small | | 0 | too thin/small for DSPC |
| blue filament | | 0 | too thin/small for DSPC |
| cyanobacteria colony | cyanobacteria colony | 1 | |
| aphanizomenon flos-aquae | aphanizomenon flos-aquae | 1 | we see it because it's mostly forming mats |
| aphanocapsa sp. | cyanobacteria colony | 1 | |
| aphanothece sp. | cyanobacteria colony | 1 | |
| chroococcus sp. | chroococcus sp. | 1 | |
| coelosphaerium sp. | coelosphaerium sp. | 1 | |
| chroococcales diverse | cyanophyceae | 1 | |
| dolichospermum sp. | dolichospermum sp. | 1 | |
| leptothrix echinata | | 0 | too thin/small for DSPC |
| leptothrix ochracea | | 0 | too thin/small for DSPC |
| merismopedia sp. | merismopedia sp. | 1 | |
| microcystis wesenbergii | microcystis sp. | 1 | |
| microcystis aeruginosa | microcystis sp. | 1 | |
| microcystis sp. | microcystis sp. | 1 | |
| planktohtrix sp. | planktohtrix sp. | 1 | |
| pseudoanabena sp. | | 0 | too thin/small for DSPC |
| phormidium | phormidium sp | 1 | |
| snowella lacrustris | snowella lacrustris | 1 | |
| spirulina sp. | | 0 | too thin/small for DSPC |
| synechococcus | | 0 | |
| woronichinia naegeliana | woronichinia naegeliana | 1 | old name: Gomphosphaeria, able to see when changing light settings |
| **Gold algae** | | | |
| chrysophyceae div | chrysophyceae div | 1 | |
| bicosoeca sp. | | 0 | too thin/small for DSPC |
| bitrichia sp. | | 0 | too thin/small for DSPC |
| chrysochromulina | | 0 | too thin/small for DSPC |
| chromulina sp. | | 0 | too thin/small for DSPC |
| dinobryon bavaricum | dinobryon sp. | 1 | to be sure we identify to genus level |
| dynobrion sp. | dinobryon sp. | 1 | to be sure we identify to genus level |
| chrysoflagellaten div | | 0 | too thin/small for DSPC |
| erkenia sp. | | 0 | too thin/small for DSPC |
| kephyrion sp. | kephyrion sp. | 1 | |
| mallomonas big | mallomonas big | 1 | |

| mallomonas acaroides | mallomonas big | 1 | |
|---|---|---|---|
| mallomonas akrokomos | mallomonas akrokomos | 1 | |
| ochromonas sp. | ochromonas sp. | 1 | |
| pseudopedinella | pseudonedinella sp. | 1 | too thin/small for DSPC |
| salpingoeca sp. | | 0 | |
| uroglena | uroglena | 1 | almost too big for 5p0x magnification |
| **Diatoms** | | | |
| stephanodiscus | centrales | 1 | impossible to see differences between cyclotella and stephanodiscus with DSPC |
| stephanodiscus big | centrales | 1 | |
| asterionella formosa | asterionella formosa | 1 | |
| aulacoseira | aulacoseira | 1 | |
| cyclotella comta | centrales | 1 | |
| centrales medium | centrales | 1 | |
| centrales small | centrales | 1 | |
| diatoma | pennales | 1 | |
| fragilaria crotonensis | fragilaria sp. | 1 | |
| gomphonema | gomphonema | 1 | |
| gyrosigma | gyrosigma | 1 | |
| navicula | pennales | 1 | |
| nitzschia | pennales | 1 | |
| pennate | pennales | 1 | |
| synedra acus | synedra sp. | 1 | |
| synedra acus angustissima | synedra acus angustissima | 1 | |
| synedra cyclopum | synedra cyclopum | 1 | |
| synedra sp. | synedra sp. | 1 | |

| **Dinoflagellates** | | | |
|---|---|---|---|
| dinoflagellates div | dinoflagellates | 1 | |
| ceratium hirundinella | ceratium hirundinella | 1 | |
| gymnodinium helveticum | dinoflagellate | 1 | |
| gymnodinium lantzschii | gymnodinium lantzschii | 1 | |
| peridinium aciculiferum | dinoflagellate | 1 | |
| peridinium diverses | peridinium sp. | 1 | |
| peridinium willei | peridinium willei | 1 | |
| dinocsyten div | dinocsyten div | 1 | |
| **Chryptophyceae** | | | |
| cryptohpytes div | cryoptophyceae | 1 | |
| chroomonas sp. | | 0 | too thin/small for DSPC |
| cryptomonas 1 | cryptomonas 1 | 1 | |
| cryptomonas 2 | cryptomonas 2 | 1 | |
| cyathomonas | | 0 | too thin/small for DSPC |
| katablephris sp. | | 0 | too thin/small for DSPC |
| rhodomonas big | rhodomona sp. | 1 | |
| rhodomonas lacustris | rhodomonas lacustris | 1 | |
| **Green algae** | | | |
| ankyra ancora | ankyra sp. | 1 | too thin to see details for sp identification |
| ankyra lanceolata | ankyra sp. | 1 | too thin to see details for sp identification |
| ankyra sp. | ankyra sp. | 1 | |
| carteria sp. | chlorophyte | 1 | |
| chlamydocapsa planctonica | chlorophyte | 1 | |
| chlamydomonas sp. | chlorophyte | 1 | |

| chloro diverse small | chlorophyte | 1 | |
|---|---|---|---|
| chloro diverse | chlorophyte | 1 | |
| chloro colonial | chloro colonial | 1 | |
| chloro diverse flagellated | chlorophyte | 1 | |
| chloro star shaped sheath | chlorophyte | 1 | |
| coelastrum microporum | coelastrum microporum | 1 | |
| coelastrum reticulatum | coelastrum reticulatum | 1 | |
| crucigenia | crucigenia | 1 | |
| crucigeniella | crucigeniella | 1 | |
| elakatothrix gelatinosa | elakatothrix gelatinosa | 1 | |
| eudorina | eudorina sp. | 1 | |
| gonium sp. | gonium sp. | 1 | |
| hormidium | hormidium | 1 | |
| lagerheimia | oocystaceae | 1 | |
| monoraphidium | | 0 | too thin/small for DSPC |
| nephrocytium | nephrocytium | 1 | |
| eutetramorus | chlorophyte | 1 | |
| green colony | chlorophyte | 1 | |
| oocystis sp. | oocystaceae | 1 | |
| oocystis marsoni | oocystaceae | 1 | |
| pandorina sp. | pandorina sp. | 1 | |
| phacotus lenticularis | phacotus lenticularis | 1 | |
| pediastrum boryanum | pediastrum sp. | 1 | |
| pediastrum duplex | pediastrum sp. | 1 | |
| pediastrum simplex | pediastrum sp. | 1 | |
| planktosphaeria | planktosphaeria | 1 | |

| galatinosa | galatinosa | | |
|---|---|---|---|
| raysiella sp. | raysiella sp. | 1 | |
| scenedesmus armatus | scenedesmus sp. | 1 | |
| scenedesmus ellipticus | scenedesmus ellipticus | 1 | |
| sphaerocystis | chlorophyte | 1 | |
| tetrachlorella alterans | chlorophyte | 1 | |
| tetraedron minimum | tetraedron sp. | 1 | |
| tetraedron sp. | tetraedron sp. | 1 | |
| ulothrix | chlorophyte filament | 1 | |
| willea irregluaris | willea sp. | 1 | |
| **Zygnematales** | | | |
| zygnemophyceae | chlorophyte filament | 1 | |
| closterium acutum | closterium acutum | 1 | too difficult to see details for sp identification |
| closterium acutum var. Variabile | closterium acutum | 1 | too difficult to see details for sp identification |
| cosmarium sp. | cosmarium sp. | 1 | |
| stauratsrum sp. | stauratsrum sp. | 1 | |
| **Ciliates** | | | |
| ciliates div | ciliphora | 1 | |
| askenasia sp. | askenasia sp. | 1 | |
| ciliates big | ciliphora | 1 | |
| ciliates small | ciliphora | 1 | |
| coleps sp. | coleps sp. | 1 | |
| didinium sp. | didinium sp. | 1 | |
| epistylis sp. | epistylis sp. | 1 | |
| halteria sp. | ciliphora | 1 | |
| lagynophrya | ciliphora | 1 | |
| lionotus | ciliphora | 1 | |
| strobilidium | ciliphora | 1 | |
| strombidium viride | strombidium sp. | 1 | |
| tintinnidium fluviatile | tintinnidium sp. | 1 | |
| tintinnopsis lacustris | tintinnopsis lacustris | 1 | |
| urothricha sp. | ciliphora | 1 | |
| vorticella | vorticella | 1 | |
| zooflagellaten div | zooflagellaten div | 1 | |
| **Rotifer** | | | |
| rotatorien div | rotifer | 1 | |
| asplachna sp. | asplachna sp. | 1 | |
| brachionus sp. | brachionus sp. | 1 | |
| conochilus sp. | conochilus sp. | 1 | |
| gastropus sp. | gastropus sp. | 1 | |
| filinia terminalis | filinia sp. | 1 | |
| kellikotia sp. | kellikotia sp. | 1 | |
| keratella cochlearis | keratella cochlearis | 1 | |
| keratella quadrata | keratella quadrata | 1 | |
| polyarthra sp. | polyarthra sp. | 1 | |
| paradileptus sp. | paradileptus sp. | 1 | Never seen in traditional sampling and microscopy but very common in DSPC 0p5x magnification |
| trichocerca sp. | trichocerca sp. | 1 | |
| synchaeta sp. | synchaeta sp. | 1 | |
| **Larger zooplankton** | | | |
| bosmina | bosmina | 1 | |
| diaphanosoma | diaphanosoma | 1 | |
| daphnia sp. | daphnia sp. | 1 | |

| cyclops | cyclops | 1 | |
|---|---|---|---|
| eudiaptomus | eudiaptomus | 1 | |
| nauplia | nauplia | 1 | |
| leptodora kindtii | leptodora kindtii | 1 | |
| chaoborus | chaoborus | 1 | |
| **Other invertebrates** | | | |
| hydra sp. | hydra sp. | 1 | |

Fig. S1

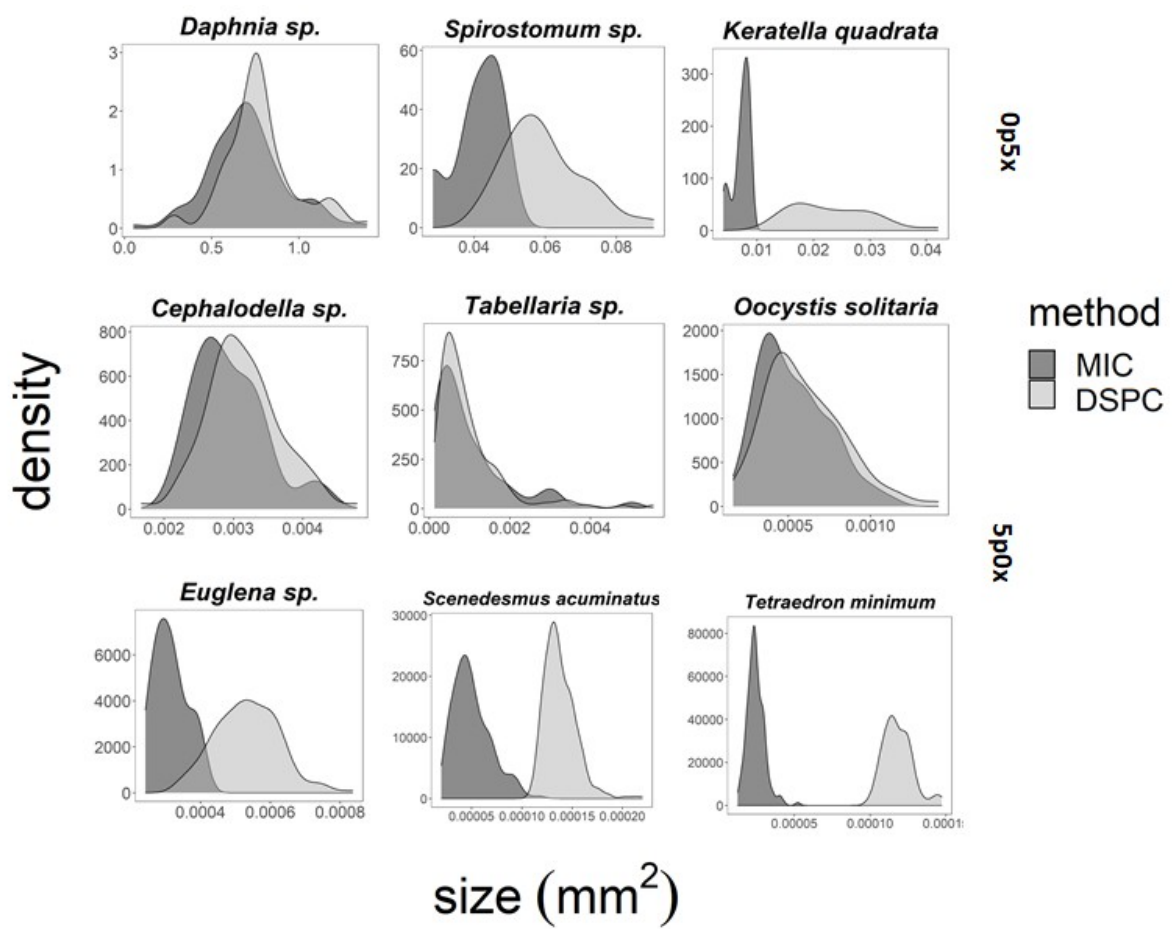

**Fig S1. Plankton body size distribution estimated by the DSPC and microscopy.** We calculated size density curves for plankton taxa covering a large size spectrum. We used

both magnifications, 0p5x and 5p0x and ordered them according to size, with the largest organisms in the left top corner and smallest in the bottom right corner. Curves are colored according to the method used, microscopy or the plankton camera. Size distributions overlap better for larger than smaller taxa in both magnifications.

Fig. S2

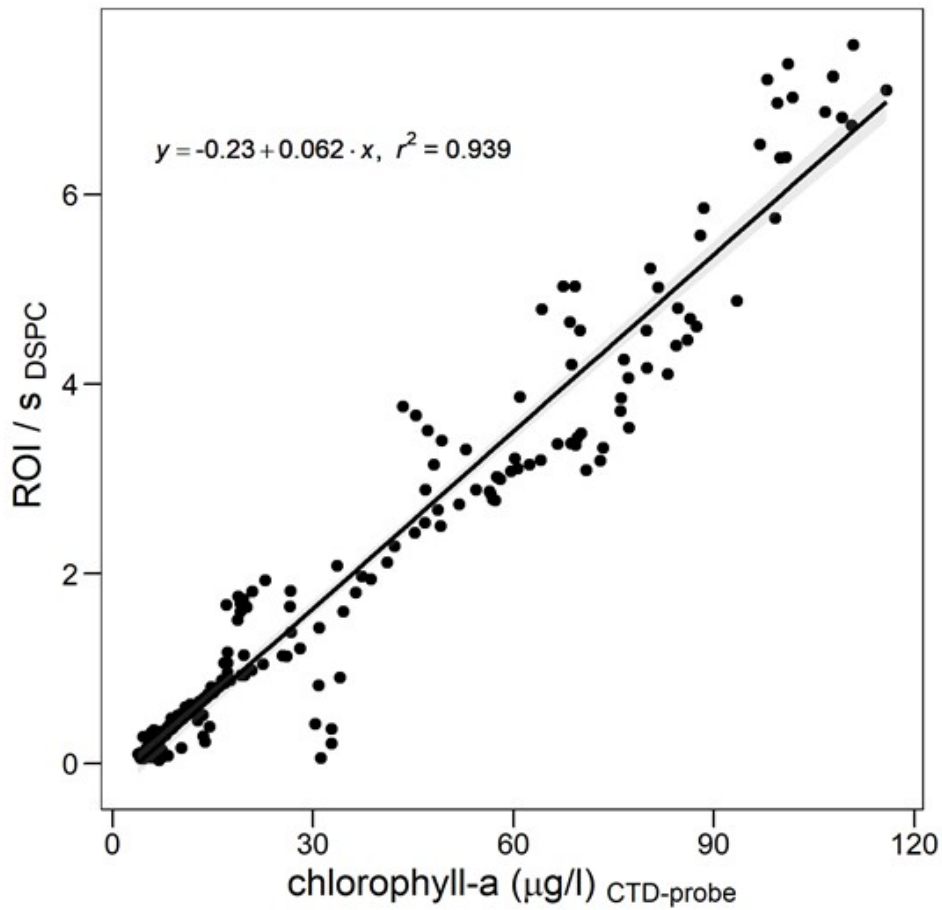

**Fig. S2. Chlorophyll-a measured by CTD against images per seconds taken by the DSPC.** Each point represents a sample taken in the year 2018. Amount of Chl-a scales well with ROI/s taken by the higher magnification (5p0x) of the plankton camera ($R^2 = 0.94$).

Fig. S3

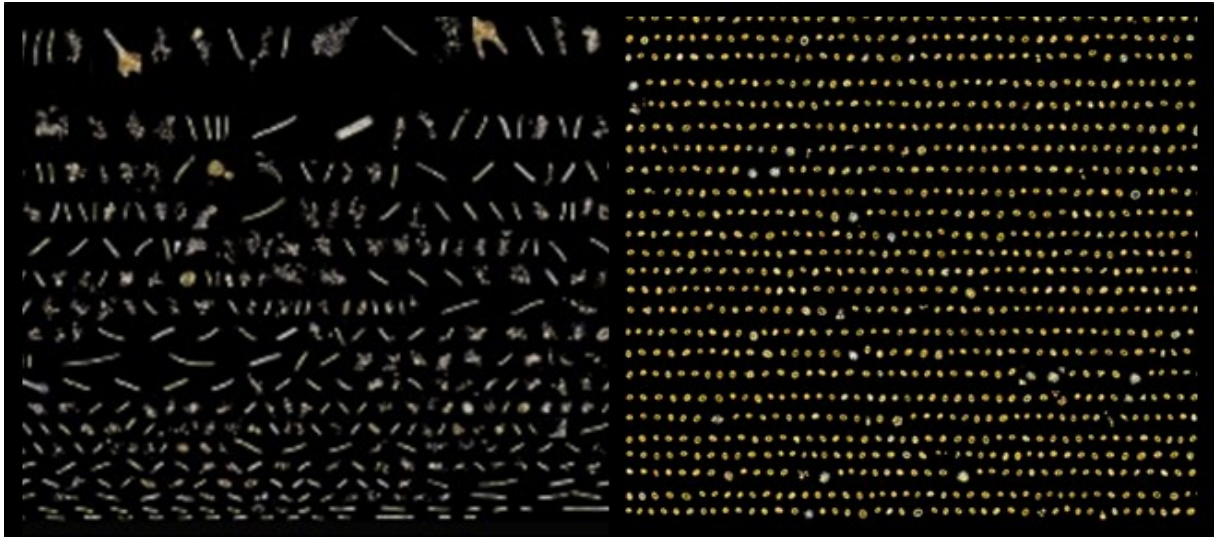

**Fig. S3. Phytoplankton community before and during the bloom of 2018 seen through the plankton camera.**  A Community before the bloom, mostly dominated by *Hormidium sp.*. B Community during the bloom, mostly dominated by *Oocystis sp.*.

# Fig. S4

**A**

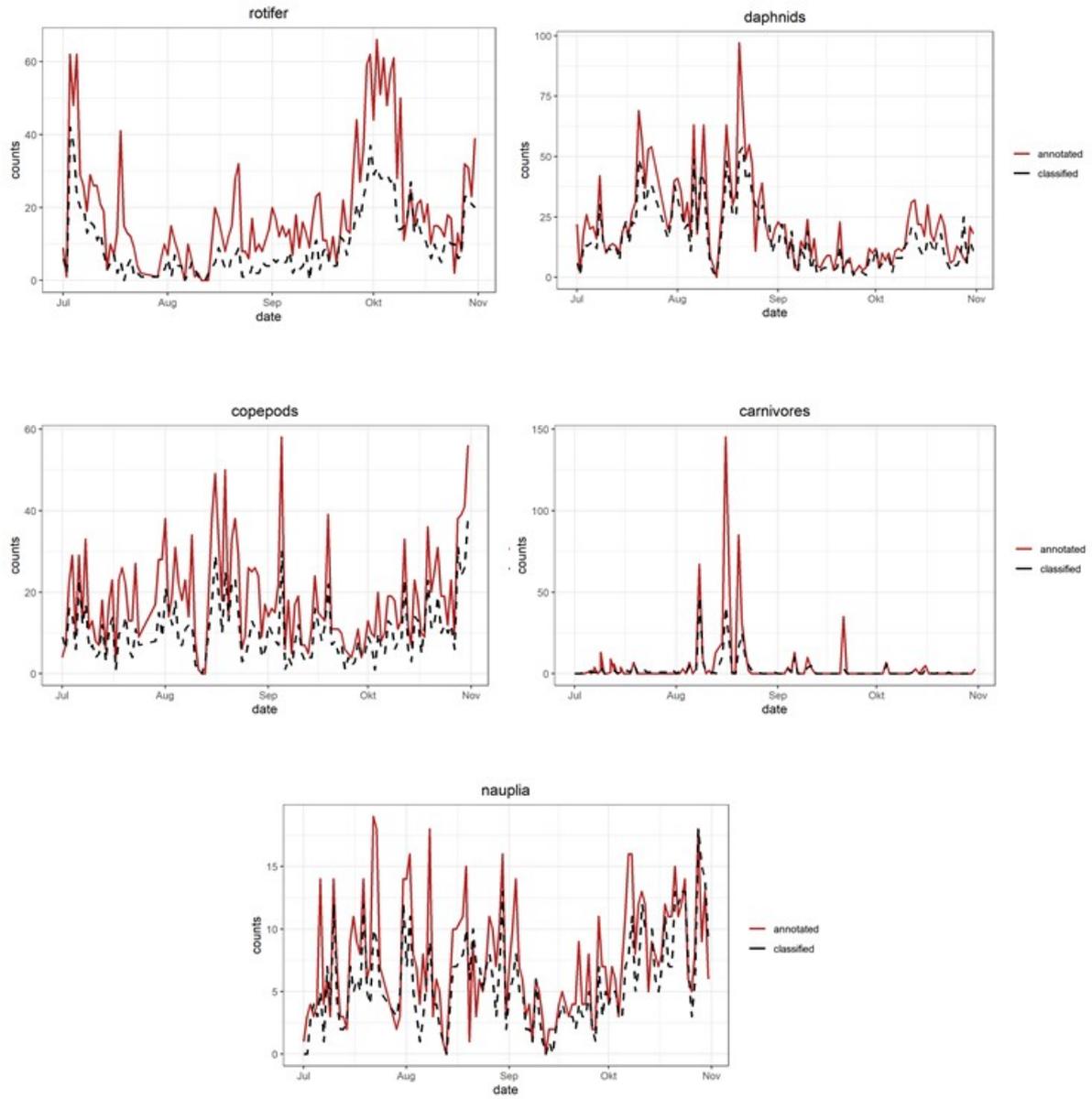

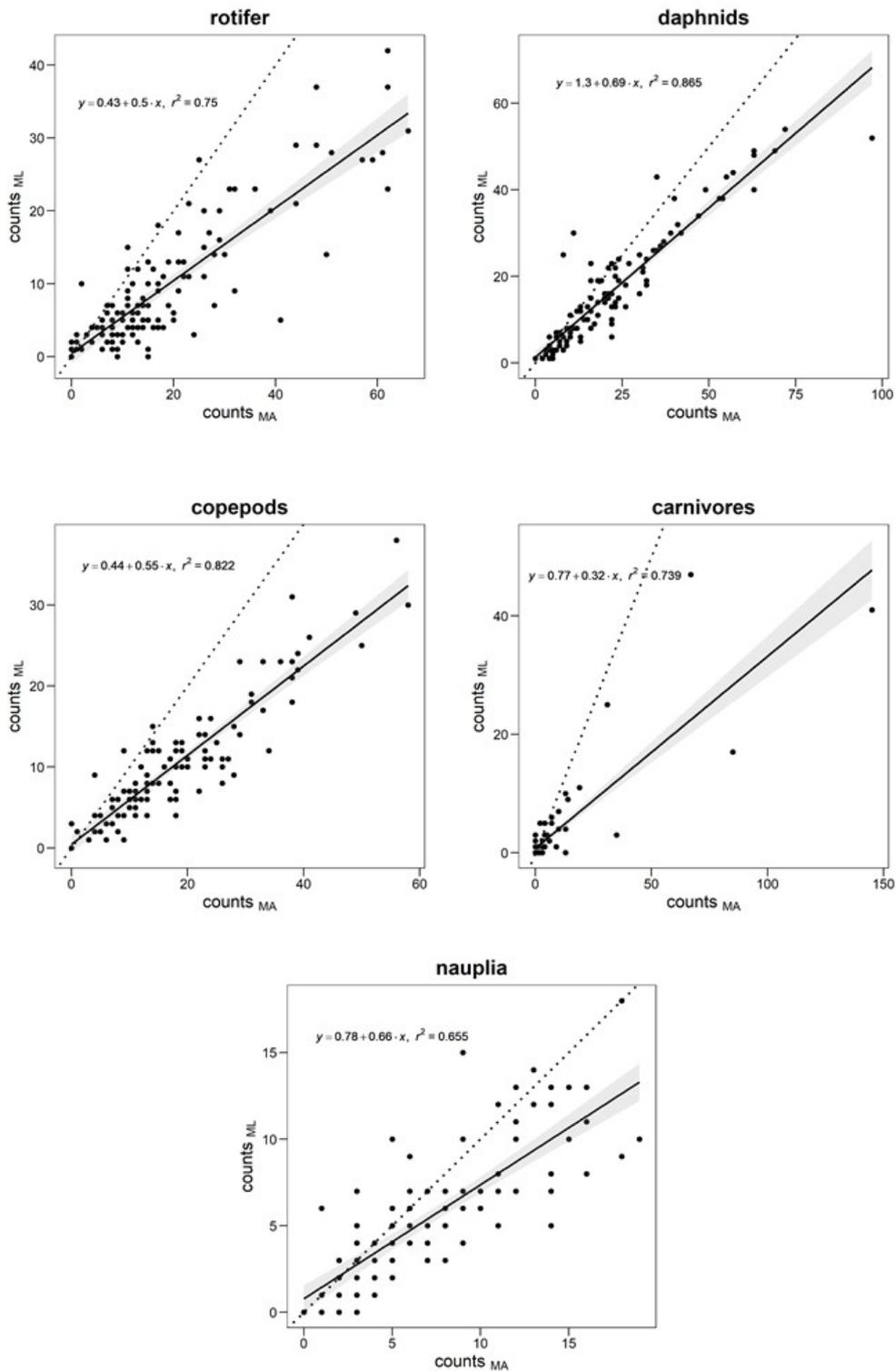

**Fig. S4: Comparison of manual annotation (MA) and classification by CCN (ML) for zooplankton taxa.** Images of 4 months of consecutive plankton camera data (one sample per day, taken during the night at 00:00 or 01:00) were manually annotated by 10 different people guided by taxonomists. The same images were run through a CNN. Zooplankton taxa were aggregated into higher taxonomic groups. **A**: Time series estimated by manual annotation (red) and CNN (black) for higher zooplankton taxa in July-November 2018, **B**:

Linear regression plots comparing manual annotation to CNN for aggregated zooplankton taxa. Each dot represents one sample between July-November 2018. Data is the same as in Fig. 5 B, seasonal patterns and changes in planktonic food web during an algal bloom.

Fig. S5

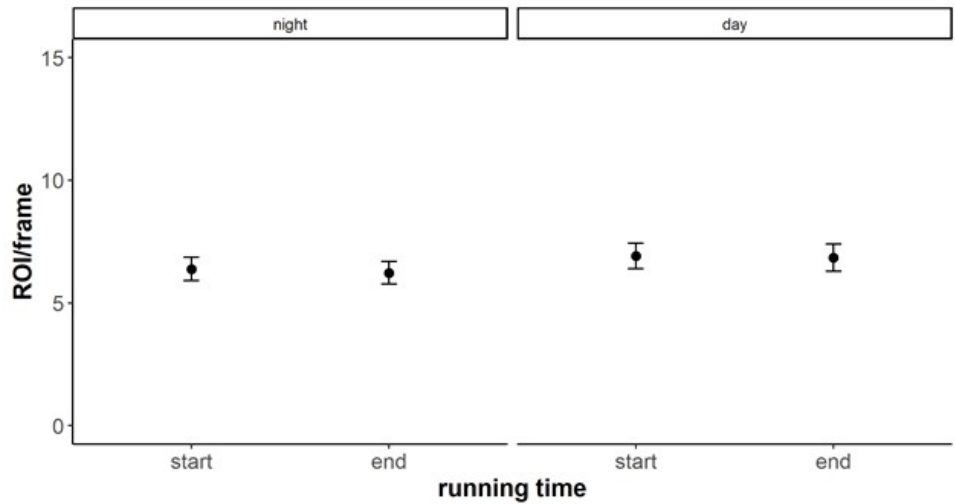

**Fig S5. Zooplankton phototaxis.** Some zooplankon species, such as *Daphina sp.*, may be attracted by the flash of the DSPC and bias results. We calculated ROI/s from the stationary DSPC in Lake Greifensee at the start and end of running time (1st minute and 9th minute). We used the whole plankton data of 2019 (n=563), March 21 - December 2019, taking a sample at 04:00 (night) and 16:00 (day) per day. Images were subset to every 6' in order to reduce duplicated ROI. There was no significant difference in ROI/s between day and night samples nor at the start and end of DSPC running time.

Fig. S6

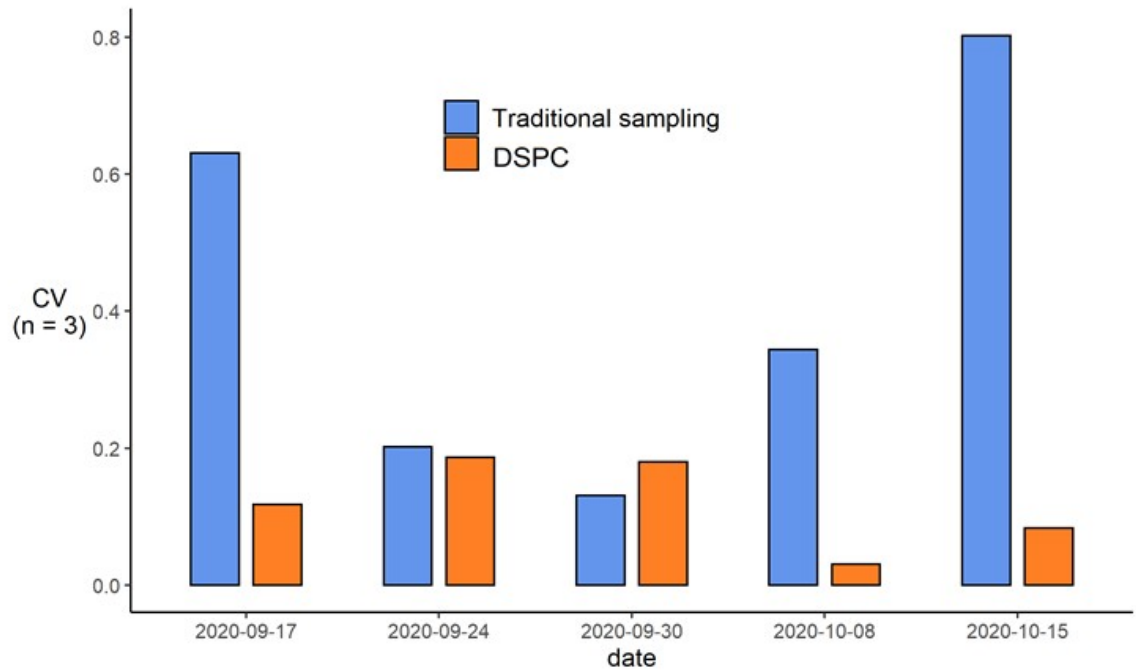

**Fig. S6. Reproducibility of abundance estimates (sampling by niskin bottle and microscopy relative to DSPC).** CV represents the coefficient of variation in zooplankton. The coefficient of variation between traditional samples was obtained from stereomicroscopy counts of three independent niskin bottles taken consequently, during routine monitoring work (middle of the morning). DSPC data come from measurements at 09:00, 10:00 and 11:00 am on the same day.

Fig. S7

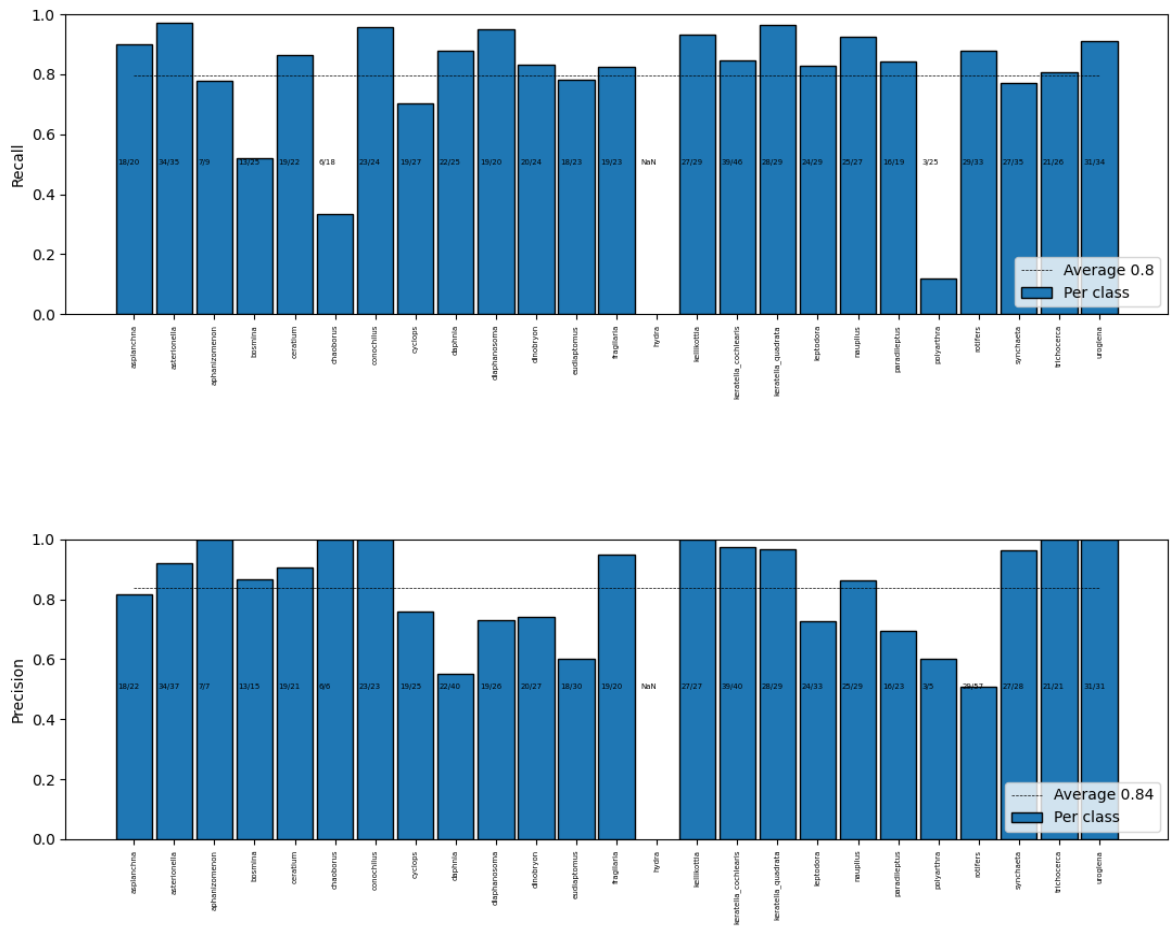

**Fig. S7. Base performance of the deep-learning classifier.** The top plot shows the per-class recall of our classifier. The bottom one shows the per class precision. The horizontal lines are the macro-average, over all the classes. The numbers X/Y on each bar indicate how many examples were correctly classified (X), out of the total (Y). This validation was performed on a completely new batch of field data, which did not contain hydra.